\documentclass[a4paper,12pt]{article}
\usepackage{bbm}

\usepackage{jheppub} 

\usepackage[T1]{fontenc} 

\usepackage[percent]{overpic}
\usepackage{wrapfig}
\usepackage{bbm}
\usepackage{tabu}
\usepackage{slashed}
\usepackage{diagbox}

\usepackage{graphicx}
\usepackage{tikz}
\usetikzlibrary{arrows,chains,shapes,matrix,positioning,scopes}
\usetikzlibrary{decorations.markings}
\tikzstyle arrowstyle=[scale=1]
\tikzstyle directed=[postaction={decorate,decoration={markings,
    mark=at position .65 with {\arrow[arrowstyle]{stealth}}}}]
\tikzstyle reverse directed=[postaction={decorate,decoration={markings,
    mark=at position .65 with {\arrowreversed[arrowstyle]{stealth};}}}]
\usepackage{import}
\usepackage{accents}
\usepackage{mathrsfs,amsmath,amssymb,slashed}
\usepackage{multirow,multicol}
\usepackage{enumitem}
\usepackage[percent]{overpic}
\usepackage{slashed}

\usepackage{wrapfig}
\usepackage{tabu}
\newcommand{\nn}{\nonumber}
\usepackage{diagbox}
\usepackage{romannum}
\usepackage{comment}
\usepackage{tikz}
\usepackage{mathrsfs,amsmath,amssymb,amsthm,amsfonts,graphicx,accents,color}
\usepackage{leftidx}
\usepackage{import}
\usetikzlibrary{decorations.pathmorphing}
\DeclareFontFamily{OT1}{rsfs}{}

\DeclareFontShape{OT1}{rsfs}{m}{n}{ <-7> rsfs5 <7-10> rsfs7 <10->rsfs10}{} 

\DeclareMathAlphabet{\mycal}{OT1}{rsfs}{m}{n}
\newcommand{\di}{\text{d}}

\newcommand{\be}{\begin{equation}}
\newcommand{\ee}{\end{equation}}

\newcommand{\F}{\mathcal{F}}
\newcommand{\A}{\mathcal{A}}
\newcommand{\kk}{\text{k}}

\newcommand{\laa}{\lambda}

\newcommand{\eqs}[1]{\begin{equation} #1 \end{equation}}

\newcommand{\D}{\mathcal{D}}
\newcommand{\M}{\mathcal{M}}

\newcommand{\C}{\mathcal{C}}

\newcommand{\p}{\prime}

\newcommand{\blist}{\begin{itemize}}

\newcommand{\ord}[1]{\mathcal{O}(#1)}

\newcommand{\un}[1]{\underline{#1}}

\DeclareMathOperator{\extdm}{d}
\newcommand{\extd}{\extdm \!}

\newcommand\enote[1]{\textcolor{red}{\bf [Erfan:\,#1]}}

\title{{\LARGE{$p$-Form Surface Charges on AdS:}\\ \large{Renormalization and Conservation}}}
\author{Erfan Esmaeili$^{a,b}$,}
\author{Vahid Hosseinzadeh$^a$}

\affiliation{\it $a$ School of Physics, Institute for Research in Fundamental
Sciences (IPM),\\  P.O.Box 19395-5531, Tehran, Iran\\
\it $b$ Erwin Shr\"odinger Institute for Physics and Mathematics\\ Boltzmanngasse 9,
1090 Wien,
Austria
}

\emailAdd{erfanili@ipm.ir}
\emailAdd{v.hosseinzadeh@ipm.ir}

\preprint{IPM/P-2021/nnn}

\abstract{Surface charges of a $p$-form theory on the boundary of an AdS$_{d+1}$ spacetime are computed. Counter-terms on the boundary produce divergent corner-terms which holographically renormalize the symplectic form. Different choices of boundary conditions lead to various expressions for the charges and the associated fluxes. With the  usual standard AdS boundary conditions, there are conserved zero-mode charges.
Moreover, we explore two leaky boundary conditions which admit an infinite number of charges forming an Abelian algebra and non-vanishing flux.
Finally, we discuss magnetic $p$-form charges and electric/magnetic duality. }    
\begin{document}
\maketitle
\newpage
\section{Introduction}
The relation between asymptotic symmetries, memory effects and soft theorems has provoked considerable progress in understanding the IR properties of field theories in the asymptotically flat spacetimes  \cite{Strominger:2014pwa,Strominger:2013lka, He:2014laa,Campiglia:2015qka}. See also \cite{Strominger:2017zoo} and the references therein.  Similar to global symmetries, large gauge transformations/diffeomorphisms lead to Ward identities in the quantum theory, as can be seen in the path integral formulation \cite{Avery:2015rga}.  Physical boundary conditions for the gauge field or the metric can be assigned such that the equivalence of Ward identities with soft theorems hold. For instance, subleading soft graviton theorems in four dimensional asymptotically flat space, suggest an extension of the standard BMS group to include superrotations \cite{Kapec:2014opa,Barnich:2009se,Campiglia:2014yka}. Identifying the correct set of boundary conditions that describe localized systems is a key step in understanding holography in asymptotically flat spacetimes, as it specifies the symmetries of the putative CFT in the semi-classical regime \cite{Barnich:2010eb,Cheung:2016iub, Ball:2019atb,Banerjee:2020kaa}.

An insightful approach to holography in asymptotically flat spacetimes is to embed these spaces in a larger Anti-de Sitter space (AdS) \cite{Polchinski:1999ry,Susskind:1998vk,Barnich:2012aw}. In particular, the relation between boundary CFT Ward identities and bulk soft theorems in the flat (i.e. large AdS radius) limit is explored in \cite{Hijano:2019qmi,Hijano:2020szl}. A natural expectation is to find appropriate boundary conditions for AdS theories which reduce to the standard asymptotically flat boundary conditions in large AdS radius limit. The standard treatment of AdS$_{d+1}$ Einstein gravity is to fix a conformal class of metrics on the boundary, for which the asymptotic symmetry group is isomorphic to the $d$ -dimensional conformal group. Otherwise, by fixing Neumann or mixed boundary conditions, the boundary metric is allowed to fluctuate which translates into energy flux on the boundary \cite{Compere:2008us,Krishnan:2016dgy}. In AdS$_4$ Comp\`ere \emph{et al} found that by choosing a foliation and a measure on the boundary manifold, the asymptotic symmetry group is the so-called $\Lambda$-BMS group \cite{Compere:2019bua,Compere:2020lrt} which reduces to the centerless BMS$_4$ in the large AdS radius limit. See also \cite{Fiorucci:2020xto} for extension to higher dimensions.



In a previous study \cite{Esmaeili:2019mbw}, we considered Maxwell theory in Anti-de Sitter space and constructed the infinite-dimensional asymptotic symmetry algebra. To analyze this problem we adopted hyperbolic slicing of the AdS space. The phase space consisted of charged particles moving freely and scattering off at the origin of the hyperbolic coordinate system. In this special system, we can treat the boundary like the spatial infinity of Minkowski space as pointed out in \cite{Hijano:2020szl} and henceforth we expect charge conservation. In this paper, we generalize the boundary conditions to include generic solutions to Maxwell's field equations including radiation that reach the boundary, as well as generalization to $p$-form theories. 

Higher form gauge fields are natural objects of study from a supergravity and string theory perspectives \cite{Horowitz:1991cd,Freedman:2012zz}. $p$-brane solutions  couple to and source the $p+1$-form gauge fields. $p$-branes and likewise $p+1$-forms, may be residing on various background geometries, such as flat or AdS backgrounds.
The main objective of this paper is a near-boundary analysis of $p$-form gauge fields and constructing the surface charges associated with residual gauge transformations.

A clear and thorough treatment of field equations and Cauchy development for scalar theory, Maxwell theory and linearized gravity in $d+1$ dimensional Anti-de Sitter space is performed by Ishibashi and Wald \cite{Ishibashi:2004wx}. For Maxwell theory, that involves a decomposition of gauge field components into scalar and vector modes with respect to rotation group. Here we first generalize their analysis to a $(p+1)$-form gauge field, for which the components can be expanded in terms of $p$-form and $p+1$-form spherical harmonics on the sphere $S^{d-1}$.  
These two sets differ in their asymptotic behavior. We discuss normalizablity of these modes in different dimensions. 

In Lorentzian AdS, non-normalizable modes are fixed as background fields upon which the normalizable modes fluctuate \cite{Balasubramanian:1998sn,Marolf:2004fy}. For some $p, d$ values both bulk modes become normalizable  \cite{Breitenlohner:1982jf}, as we specify for different components of $p$-form fields. In our case, non-normalizable modes are those which make the symplectic form divergent. However, the symplectic form is modified when counter-terms are added on the boundary. Subtracting the divergences in the on-shell action is provided by  local Lagrangians defined on the AdS boundary \cite{Balasubramanian:1999re,Skenderis:2002wp}. Variation of the counter-terms produce total divergence terms, which are known as corner terms. These divergent contributions modify the symplectic form of the theory and make it finite \cite{Papadimitriou:2005ii,Compere:2008us,Freidel:2019ohg}. We will exhibit the counter-term for $p$-form theory and construct the renormalized symplectic form, which guarantees finiteness of surface charges.

The residual gauge transformations after fixing the radial gauge have finite  surface charges that follow from the renormalized symplectic structure. We derive the charge expression without imposing boundary conditions, and in particular allow finite  gauge transformations  on the boundary. For a Maxwell field this amounts to defining  the Dirichlet condition  as 
\begin{equation}
\delta F^{(0)}|_{\mathscr{I}}=0
\end{equation}
(instead of the usual condition $\delta A^{(0)}|_{\mathscr{I}}=0$). The generalized boundary condition that leads to this kind of constructions is in general leaky: one can not demand conservation and also infinite asymptotic symmetry  at the same time. Thus, the flux on AdS boundary is non-vanishing except for the $U(1)$ electric charge. Addition of boundary terms to the action leads to different charge expressions and different fluxes with no preferred criterion for choosing one.   We examine two cases of boundary conditions with flux and derive the charge expressions for the boundary gauge mode (sketched in table \ref{Summary table}).

As it turns out, conservation of the charges is closely related to the existence of an energy functional on the solution space. The reason is that addition of the boundary gauge transformations deforms the expression for the energy by a surface term and spoils its integrability. We discuss specific configurations with conserved charges and an integrable energy functional.

This paper is organized as follows. In section \ref{Wald}, the asymptotic behavior of $p$-form fields and their norms on Anti-de Sitter space is discussed. In section \ref{Renormalization section}, we explain the renormalization procedure for the symplectic form upon which the later discussion is built. In section \ref{Charge section}, we demonstrate how different charge expressions fail to be conserved and identify the conservative backgrounds.  Section \ref{EM duality section} is devoted to electric/magnetic duality as  a symmetry of the $p$-form action and its relation to boundary conditions and magnetic charges (see also \cite{DeWolfe:2020uzb}). Finally, section \ref{energy} discusses non-integrability of energy on the phase space.

\paragraph{Conventions and Notation.} We will use capital Latin indices  $I,J,K$ for denoting the coordinates of the AdS$_{d+1}$ space. We decompose the spacetime to 2-dimensional radial and time direction denoted by $x^a \equiv \{t,r\}$ and the rest $n=d-1$-dimensional sphere denoted by $z^i$. The coordinates on the boundary is also denoted by $x^{\mu}\equiv \{t,z^i\}$.

\section{The asymptotic behaviour of gauge field in AdS, normalizable modes}\label{Wald}
In this section, we review the asymptotic behaviour of $p+1$-from gauge field on AdS$_{d+1}$ based on the Ishibashi-Wald work \cite{Ishibashi:2004wx}. We start by decomposition of the gauge field in terms of exact and co-exact forms on the sphere and define two gauge invariant functions out of them. Then we write the symplectic from in terms of these functions. At last, we identify the  normalizable modes defined as modes that keep the symplectic form finite. 
\subsection{Falloff conditions and equations of motions}
The action is 
\begin{align}\label{theory}
    S=\frac{-1}{2(p+2)!}\int_{\mathcal{M}} \sqrt{g}\F_{ \un{I}^{p+2}}\F^{ \un{I}^{p+2}}\,,\qquad \F_{ \un{I}^{p+2}}=(p+2)\partial_{[I_1}\A_{I_2\cdots I_{p+2}]}
\end{align}
where $\mathcal{M}$ is AdS spacetime and we  work in global coordinates
\begin{align}
    ds^2=&\frac{1}{\sin^{2}x}\big[-dt^2+dx^2+\cos^2x d\Omega_{d-1}^2\big]
    =-(r^2+1)dt^2+\frac{dr^2}{r^2+1}+r^2 q_{ij}(z) dz^i dz^j
\end{align}
where $r=\cot{x}$ and the conformal boundary is at $x=0$. $q_{ij}(z)$ is the round unit sphere metric on of $S^n$.  The metric on the conformal boundary, $R \times S^n$ is set as
\begin{align}
    h_{\mu\nu}dx^\mu dx^\nu=-dt^2+q_{ij}(z)dz^i dz^j.
\end{align}
Given a radial coordinate $r$ in AdS, The equation of motion $\nabla_I F^{I\un{J}^{p+1}}=0$   breaks into two sets
\begin{align}
    \nabla_\mu \F^{\mu r\un{\nu}^{p}}=0  \qquad \qquad \text{Constraints},\label{Ceq}\\
    \nabla_I \F^{I\un{\mu}^{p+1}}=0 \qquad \qquad \text{Dynamical}.\label{Deq}
\end{align}
Eq. \eqref{Ceq} can be interpreted as the constraints on the boundary of AdS and eq. \eqref{Deq}  determines the bulk fields in terms of their values on the boundary. The existence of the constraints on the boundary is due to the gauge structure of the theory.

We exhibit field equation in terms of gauge invariant variables made out of the gauge field and its derivatives, following the analysis of Ishibashi and Wald \cite{Ishibashi:2004wx}. For this purpose it is convenient to define an $AdS_2$ metric $g_{ab}$ to write the spacetime metric as
\begin{equation}
    ds^2=g_{ab}(x)dy^a dy^b+r^2 q_{ij}(z) dz^i dz^j\qquad0<x\leq\frac{\pi}{2}.
\end{equation}
We start by decomposing different components of the gauge field in terms of exact and co-exact forms on the $n$-sphere. 
\begin{subequations}\label{T decompose}
\begin{align}
    \A_{\un{i}^{p+1}}&=\sum_k\Big(\psi^{(k)}\mathbb{T}^{(k)}_{\un{i}^{p+1}}+f^{(k)}d\mathbb{T}^{(k)}_{\un{i}^{p}}\Big)\\
     \A_{a\un{i}^{p}}&=\sum_k\Big(f^{(k)}_a\mathbb{T}^{(k)}_{\un{i}^{p}}+g_a^{(k)}d\mathbb{T}^{(k)}_{\un{i}^{p-1}}\Big)\\
      \A_{ab\un{i}^{p-1}}&=\sum_k\Big(f^{(k)}_{ab}\mathbb{T}^{(k)}_{\un{i}^{p-1}}+g_{ab}^{(k)}d\mathbb{T}^{(k)}_{\un{i}^{p-2}}\Big)
\end{align}
\end{subequations}
where $\mathbb{T}^{(k)}_{\un{i}^{p}}$ is a $p$-form spherical harmonic on the $S^n$ and $d$ is the exterior derivative on the sphere. Note that $\psi^{(k)}$ exists when $n>p+1$. A $p$-form spherical harmonic is a divergenceless $p$-form on $n$-sphere $\mathcal{D}^j\mathbb{T}_{ji_2\cdots i_p}=0$ that diagonalizes the   Laplace-Beltrami operator $\Delta=d d^\dagger+d^\dagger d$
\footnote{$d^\dagger=\star d\star$ is the adjoint of exterior derivative $d$ with respect to the standard inner product of $p$-forms on sphere
\begin{equation}
   \langle d\alpha,\beta\rangle=\langle\alpha,d^\dagger\beta\rangle\,.
\end{equation}
}
    \begin{equation}
  \Delta \mathbb{T}^{(k)}_{\un{i}^q}=\kk^2_q\mathbb{T}^{(k)}_{\un{i}^q},\qquad\qquad  \text{k}^2_q=(k(k+d-2)+q(d-q-2))
\end{equation}
Harmonics with different $k$ are orthogonal. To fix the normalization, let us revert the implicit index $m$ that labels different $p$-form spherical harmonics $\mathbb{T}^{(k,m)}$ with the same  eigenvalue $\kk_q^2$, and demand that
\begin{equation}\label{p normalization}
    \int_{S^n}\sqrt{q}\,\mathbb{T}_{(k,m)}^{\un{i}^p}\mathbb{T}^{(k^\p,m^\p)}_{\un{i}^p}=p!\delta_{k,k^\p}\delta_{m,m^\p}
\end{equation}
We will generally suppress the index $m$ to reduce clutter.

Note that $n>p$. The field strength tensor is
\begin{subequations}
\begin{align}
    \F_{ab\un{i}^p}&=\widehat{\nabla}_a\A_{b\un{i}^{p}}-\widehat{\nabla}_b\A_{a\un{i}^{p}}+d \A_{ab\un{i}^{p-1}}\\
    \F_{a\un{i}^{p+1}}&=\widehat{\nabla}_a\A_{\un{i}^{p+1}}- d\A_{a\un{i}^{p}}\\
    \F_{\un{i}^{p+2}}&= d\A_{\un{i}^{p+1}}
\end{align}
\end{subequations}
where $\widehat{\nabla}_a$ is the covariant derivative of the metric $g_{ab}$ on AdS$_2$.
We call the coefficients of $\mathbb{T}_{\un{i}^p}$, \textbf{$p$-form components} and the coefficients of $\mathbb{T}_{\un{i}^{p+1}}$,  \textbf{$(p+1)$-form components}. $(p+1)$-form components  drop from the constraint equations. Normally, the radial gauge is fixed. In contrast, different $p$-form components appearing in the gauge field are coupled by the constraint equations to give one independent $p$-form on the boundary after fixing the radial gauge $\A_{r\un{\mu}^p}=0$. Now we solve field equations for $p$-form and $(p+1)$-form components separately.

\subsection*{$\boldsymbol{(p+1)}$-form components}
There exists only one $(p+1)$ component in the gauge field which is $\psi$. For this component we have
\begin{align}
    \F_{ab\un{i}^p}=0,\quad\qquad
    \F_{a\un{i}^{p+1}}=\sum_k\widehat{\nabla}_a\psi^{(k)}\mathbb{T}^{(k)}_{\un{i}^{p+1}},
   \qquad\quad
    \F_{\un{i}^{p+2}}= \sum_k\psi^{(k)}d\mathbb{T}^{(k)}_{\un{i}^{p+1}}\,,
\end{align}
 and the equations of motion take the form
\begin{equation}\label{psieq}
    r^{6+2p-d}\widehat{\nabla}_a\Big(r^{d-4-2p}\widehat{\nabla}^{a}\psi^{(k)}\Big)-\kk^2_{p+1}\psi^{(k)}=0.
\end{equation}
Equation \eqref{psieq} expands to
\begin{align}\label{psieom0}
  \frac{\partial^2}{\partial t^2} \psi^{(k)}= \frac{1}{r^{d-3-2p}}\frac{\partial}{\partial x}\big(r^{d-3-2p} \frac{\partial}{\partial x}\psi^{(k)})-\frac{(1+r^2)\kk_{p+1}^2}{r^2}\psi^{(k)}.
\end{align}
By re-scaling $\boldsymbol{\psi}^{(k)}= r^{(d-2p-3)/2}\psi^{(k)}$  the differential equation simplifies to
\begin{align}\label{psieqx}
    \frac{\partial^2}{\partial t^2} \boldsymbol{\psi}^{(k)}=\left(\frac{\partial^2}{\partial x^2}-\frac{\nu_1^2-1/4}{\sin^2{x}}-\frac{\sigma^2-1/4}{\cos^2{x}} \right)\boldsymbol{\psi}^{(k)}
\end{align}
with
\begin{equation}\label{nu_1 sigma}
    \nu_1=\frac{1}{2}|n_p-1|\,, \qquad\qquad \sigma=k+\frac{d-2}{2}.
\end{equation}
where we define $n_p=n-2p$.  In the form \eqref{psieqx}, the number of normalizable modes is simple to decide (basically, depending on the sign of $\nu_1-1$). The special cases $p=0$ (vector part of Maxwell theory) and $p=-1$ (scalar field) are previously studied \cite{Ishibashi:2004wx}.

We are mainly concerned with the asymptotic behavior of the solutions.  Assuming a fall-off of the form $\psi^{(k)}\sim r^{\lambda_{\psi}}$, \eqref{psieom0} implies
\begin{align}
    &(n_p-1+\lambda_{\psi})\lambda_{\psi}=0.\label{pfalloffeq}
\end{align}
The two solutions for $\lambda_\psi=0$ and $\lambda_\psi=1-n_p$. When $n_p=1$ however we must add a logarithmic term at the leading order.

\subsection*{$\boldsymbol{p}$-form components}
The non-vanishing components of the field strength arising from $p$-form harmonics are
\begin{subequations}
\begin{align}
    \F_{ab\un{i}^p}&=\sum_k\Big(\nabla_af^{(k)}_b-\nabla_bf^{(k)}_a\Big)\mathbb{T}^{(k)}_{\un{i}^{p}}\\
    \F_{a\un{i}^{p+1}}&=\sum_k\Big(\nabla_af^{(k)}-f_a^{(k)}\Big)d\mathbb{T}^{(k)}_{\un{i}^{p}}\\
    \F_{\un{i}^{p+2}}&= 0.
\end{align}
\end{subequations}
From the $\un{i}^{p+1}$ components of equation of motions $\nabla_I \F^{I\un{i}^{p+1}}=0$ and from the fact that $\F_{\un{i}^{p+2}}=0$ for the $p$-component case, we have $\widehat{\nabla}_a(r^{d-1}\F^{a\un{i}^{p+1}})=0$. This leads us to define the scalars $\phi^{(k)}(x^a)$,
\begin{equation}\label{F_ai to phi}
   \F_{a\un{i}^{p+1}}=r^{2p+3-d}\sum_k\epsilon_{ab}\widehat{\nabla}^b\phi^{(k)}d\mathbb{T}^{(k)}_{\un{i}^{p}}
\end{equation}
up to a constant shift $\phi^{(k)} \to \phi^{(k)}+ \text{const.}$ 
So we have $ r^{2p+3-d}\epsilon_{ab}\widehat{\nabla}^b\phi^{(k)}=\widehat{\nabla}_af^{(k)}-f_a^{(k)}$, which implies that $2\widehat{\nabla}_{[a}f_{b]}=\epsilon_{ab}\widehat{\nabla}_c(r^{2p+3-d}\widehat{\nabla}^c\phi)$ and we can read $\F_{ab\un{i}^p}$ in terms of $\phi^{(k)}$, 
\begin{equation}
     \F_{ab\un{i}^p}=\epsilon_{ab}\sum_k\widehat{\nabla}_c\big(r^{2p+3-d}\widehat{\nabla}^c\phi^{(k)}\big)\mathbb{T}^{(k)}_{\un{i}^{p}}.
\end{equation}
From the ${a\un{i}^{p}}$ components of \eqref{Deq} we have, 
\begin{equation}\label{EOM-ai}
  \widehat{\nabla}_b\Big(r^{d-2p-1}\widehat{\nabla}_a(r^{2p+3-d}\widehat{\nabla}^a\phi^{(k)}) -\kk_p^2 \phi^{(k)} \Big)=0.
\end{equation}
 The equation then takes the form
\begin{equation}\label{phi equation 24}
  r^{d-2p-3}\widehat{\nabla}_a(r^{2p+3-d}\widehat{\nabla}^a\phi^{(k)}) -\phi^{(k)}\frac{\kk_p^2}{r^2}=\frac{c}{r^2}.
\end{equation}
Unless $\kk_p^2=0$, the constant $c$ can be absorbed in $\phi$ by constant shift\footnote{If $\kk_p^2=0$, the solutions is $\F_{ab\un{i}^p} \approx \epsilon_{ab}cr^{-n_p}\mathbb{T}^{(0)}_{\un{i}^{p}}.$} and one obtains the useful relation 
\begin{equation}\label{tr phi}
    \F_{ab\un{i}^p} \approx \epsilon_{ab}\sum k_p^2\phi^{(k)}r^{-n_p}\mathbb{T}^{(k)}_{\un{i}^{p}}.
\end{equation}
Rewriting \eqref{phi equation 24} as,
\begin{equation}\label{eom1}
\frac{\partial^2}{\partial t^2}\phi^{(k)}=\frac{1}{r^{2p+3-d}}\frac{\partial}{\partial x}(r^{2p+3-d}\frac{\partial}{\partial x}\phi^{(k)}) - \frac{\kk_p^2(1+r^2)}{r^2} \phi^{(k)}
\end{equation}
and rescaling the field as $\boldsymbol{\phi}^{(k)}= r^{(3+2p-d)/2}\phi^{(k)}$ we arrive at the final form
\begin{align}\label{phieomm}
    \frac{\partial^2}{\partial t^2} \boldsymbol{\phi}^{(k)}=\left(\frac{\partial^2}{\partial x^2}-\frac{\nu_0^2-1/4}{\sin^2{x}}-\frac{\sigma^2-1/4}{\cos^2{x}} \right)\boldsymbol{\phi}^{(k)}
\end{align}
 where
 $  \nu_0=\frac12|d-2p-4|$
and $\sigma $ is given in \eqref{nu_1 sigma}. The particular case $p=0$ (scalar part of Maxwell theory) was previously considered  \cite{Ishibashi:2004wx}. By expanding \eqref{eom1} near the boundary and assuming the behavior $\phi^{(k)}\sim r^{\lambda_{\phi}}$ we find
\begin{align}
    (3-n_p+\lambda_{\phi})\lambda_{\phi}=0\label{p+1falloffeq}.
\end{align}
As a result, we have two independent solutions with $r^0$ and $r^{n_p-3}$ falloff behaviour at infinity. Again when $n_p=3$ we need a logarithmic term at the leading order to have two independent solutions.

\subsection*{$\boldsymbol{(p-1)}$- and $\boldsymbol{(p-2)}$-form components}
Assuming that only the coefficients of $\mathbb{T}^{(k)}_{\un{i}^{p-1}}$ and $\mathbb{T}^{(k)}_{\un{i}^{p-2}}$ in \eqref{T decompose} are non-vanishing,  the  the field strength tensor is
\begin{align}
   \F_{a\un{i}^{p+1}}=0,\qquad\qquad  \F_{\un{i}^{p+2}}= 0,\qquad \qquad \F_{ab\un{i}^p}=\sum_k\Big(\nabla_ag^{(k)}_b-\nabla_bg^{(k)}_a+f^{(k)}_{ab}\Big)d\mathbb{T}^{(k)}_{\un{i}^{p-1}}.
\end{align}
The equation of motion $\D_j \F^{jab \un{i}^{p-1}}=0$ implies $\F_{ab\un{i}^p}=0$ for $(p-1)$-form sector. This is because there is no zero mode for the Laplace-Beltrami operator on $p$ forms on $S^n$ topology if $p\neq 0,n$. Thus, $p-1$ forms do not appear in the on-shell field strength and we further conclude  that
  $  -\nabla_ag^{(k)}_b+\nabla_bg^{(k)}_a=f^{(k)}_{ab}.$ 
For the $(p-2)$-form components, all the components of the field strength are identically vanishing.

\paragraph{Degrees of freedom}The $(p+1)$-form gauge theory on $AdS_{d+1}$ has $\binom{d-1}{p+1}$ degrees of freedom \cite{Afshar:2018apx}. This can be seen by enumerating the constraints in the Hamiltonian formulation. The $(p+1)$-form components are arbitrary divergenceless $(p+1)$-forms on the $d-1$-sphere, which we have expanded on the basis $\mathbb{T}_{\un{i}^{p+1}}$. The total components of $\mathbb{T}_{\un{i}^{p+1}}$ is $\binom{d-1}{p+1}$. The divergence-free condition $\mathcal{D}^k \mathbb{T}_{k\un{i}^{p}}$ is a set of $\binom{d-2}{p}$ constraints. However, these are not independent as $\mathcal{D}^k \mathcal{D}^j \mathbb{T}_{kj\un{i}^{p-1}}=0$ identically. This reduces the number of constraints by $\binom{d-2}{p-1}$. This counting iterates as
\begin{equation}
    \binom{d-1}{p+1}-\binom{d-2}{p}+\binom{d-2}{p-1}-\cdots +(-1)^{p+1}\binom{d-1}{0}=\binom{d-2}{p+1}.
\end{equation}
Therefore, there are $\binom{d-2}{p+1}$ degrees of freedom in $p+1$-form components ($\psi$). Similar counting yields $\binom{d-2}{p}$ degrees of freedom for $p$-form components ($\phi$). For instance, 2-form theory on $AdS_5$ has two 2-form components and one 1-form component. The two numbers coincide only if $d=2p+3$ (i.e. $n_p=2$).

\subsection{Normalizability} \label{NORSection}
There are different norms for field theories on AdS. The field configurations are \emph{normalizable} if they  have finite a norm, and \emph{non-normalizable} otherwise. Normalizable modes have finite conserved energy as well. We define normalizablity with the condition that the symplectic structure
 be finite with the asymptotic behaviour at hand.

In the radial gauge, the symplectic form is, 
\begin{align}\label{original symplectic form}
    \Omega=\int_{\Sigma}  \omega^t=\frac{1}{(p+1)!}\int_{\Sigma} \sqrt{g}    \delta\A_{\un{i}^{p+1}}\delta \F^{t\un{i}^{p+1}} =\frac{1}{(p+1)!}\int_{\Sigma} \sqrt{q}  \frac{r^{n_p-2}}{r^2+1} \delta \F_{t\un{i}^{p+1}} \delta\A_{\un{i}^{p+1}}
\end{align}
where $\omega^\mu=\delta\theta^\mu$ and $\theta^\mu$ is the symplectic potential of the theory\footnote{In our convention, the $\sqrt{g}$ `factor' is absorbed in the definition of $\theta^\mu$ and $\omega^\mu$ to reduce clutter.}.
In terms of gauge invariant variables $\psi$ and $\phi$ the symplectic form is
\begin{align}
    \Omega= \Omega^{(p)}+\Omega^{(p+1)}
\end{align}
where 
\begin{subequations}\label{sympformphipsi}
\begin{align}
    \Omega^{(p)}&={\kk_p^2}\int \frac{dr}{(r^2+1)}  \delta \dot{\boldsymbol{\phi}}_{(k)}\delta \boldsymbol{\phi}^{(k)} + {\kk_p^2} \Big(\delta \phi^{(k)}  \delta f_{(k)}\Big)|_{\mathscr{I}}\\
 \Omega^{(p+1)}&=\int \frac{dr}{(r^2+1)}  \delta \dot{\boldsymbol{\psi}}_{(k)} \delta \boldsymbol{\psi}^{(k)}
\end{align}
\end{subequations}
where the sum over $k$ is understood. Prior to gauge fixing, $f_{(k)}$ can be shifted by arbitrary gauge transformations.  Note that the appearance of the gauge mode is in the $p$-form sector and the $(p+1)$-form sector is free of such a mode. We will come back to this gauge part but we  dismiss it for the moment.  If $\nu\geq 1$, the fall-offs for $\boldsymbol{\phi}$ and $\boldsymbol{\psi}$  readily imply that the symplectic form is divergent for variations coming from the leading solution. In contrast, both solutions are normalizable for for $\nu=0, 1/2$. The number of normalized modes in terms of $n_p$ for the $p$-form and $p+1$-form sectors are shown in the table \ref{NNMs}.  

One can also consider the symplectic flux in terms of $\psi$ and $\phi$, 
\begin{subequations}
\begin{align}\label{sympphipsi}
    \Omega^{(p)}_{flux}&= {\kk_p^2}\int_{\mathscr{J}} dt \, r^2 \delta \boldsymbol{\phi}^{(m)} \partial_r \delta \boldsymbol{\phi}_{(m)}+{\kk_p^2} \int_{\mathscr{J}} dt \partial_t\Big(\delta  f_{(m)} \delta \phi^{(m)}\Big)\\
   \Omega^{(p+1)}_{flux}&=\int_{\mathscr{J}} dt \, r^2 \delta \boldsymbol{\psi}^{(n)} \partial_r \delta  \boldsymbol{\psi}_{(n)}
\end{align}
\end{subequations}
Once again we see that for $\nu=0, 1/2$ the both sectors have  vanishing flux independently. For $\nu \geq 1$ however the leading solution has symplectic flux and we have just one mode with vanishing flux. This gauge invariant formulation makes it clear that  symplectic flux or symplectic convergence lead to same results for the number of normalizable mode. 

\begin{table}[h]
\centering
    \begin{tabular}{|c|c|c|}
    \hline
  $n_p$ &  $p$ sector & $p+1$ sector \\    \hline
   $<0$  &$1$ &$1$\\    \hline
   $0$  &$1$ &$2$\\    \hline
   $1$  &$1$ &$2$\\    \hline
   $2$  &$2$ &$2$\\    \hline
   $3$  &$2$ &$1$\\    \hline
   $4$  &$2$ &$1$\\    \hline
   $>4$ &$1$ &$1$\\    \hline
    \end{tabular}
    \caption{The number of independent normalizable modes for $p$ and $p+1$ form sector. $n_p = d-2p-1$. }
    \label{NNMs}
\end{table}
\subsection*{Falloff conditions in terms of the gauge field}
We conclude this section by reviewing the falloff conditions for the gauge field itself. 
For simplicity we fix the radial gauge $\A_{r\un{\mu}^{p}}=0$. The falloff conditions for the rest of the components $\A_{\un{\mu}^{p+1}}$ are obtained from those for $\phi$ and $\psi$ as 
\begin{align}\label{general falloffs}
\begin{array}{lll}
\A_{\un{\mu}^{p+1}}&=A_{\un{\mu}^{p+1}}^{[\alpha]}r^{0}+\cdots+A_{\un{\mu}^{p+1}}^{[\beta]}r^{1-n_p}+\cdots \qquad \qquad &n_p\neq 1\\
\A_{\un{\mu}^{p+1}}&=A_{\un{\mu}^{p+1}}^{[\alpha]}r^{0}+\cdots+A_{\un{\mu}^{p+1}}^{[\beta]}\ln{r}+\cdots \qquad \qquad &n_p=1
\end{array}
\end{align}
where $\A^{[\alpha]}_{\un{\mu}^{p+1}}$ and $\A^{[\beta]}_{\un{\mu}^{p+1}}$ are the two independent solutions and note that the $[\alpha]$ sector is the leading one when $n_p \geq 2$ and $[\beta]$ solution is leading for other cases.

\section{The renormalized action and  symplectic form}\label{Renormalization section}
Before we delve into renormalization, we shortly review the Harlow-Wu method for defining the symplectic form in the presence of corner terms \cite{Harlow:2019yfa}. Then we introduce the counter-terms that make the on-shell action finite. In appendix \ref{CTAppen}, we work out the divergences in the action and the symplectic form explicitly in a few examples.

\subsection{Symplectic form in presence of corner terms}  
Considering  a boundary action $S_b=\int_{\mathscr{I}} L_b$, the boundary term from variation of the total action becomes
\begin{align}
    \delta (S+S_b)\approx \int_{\mathscr{I}} \left(\theta^r+\delta L_b\right).
\end{align}
Assume that for a particular boundary condition, 
\begin{align}\label{BC-HW}
    (\theta^r+\delta L_b)|_{\mathscr{I}}=\partial_\mu \C^\mu
\end{align}
where $\C^\mu$ is a phase space one-form leading to a two-form $\omega^\mu_\C=\delta \C^\mu$. According to the  Harlow-Wu prescription, the symplectic form of the theory is
\begin{align}\label{symp C}
    \Omega=\int_\Sigma  \omega^t+\oint_{\partial\Sigma} \omega^t_\C
\end{align}
This expression is manifestly conserved, as the  symplectic flux  is  a total divergence  by the boundary condition \eqref{BC-HW}, and it is subtracted as two corner terms by the Stokes theorem. Note also that with the boundary condition \eqref{BC-HW}, the action principle is satisfied by fixing $C^\mu$ at initial and final surfaces. In addition, this definition is free from $Y$ ambiguity.

Having said that, there are  situations  that we have the following boundary condition 
\begin{align}\label{BC-HW-F}
    (\theta^r+\delta L_b)|_{\mathscr{I}}=\partial_\mu \C^\mu+\mathbbmss{F} 
\end{align}
where  $\mathbbmss{F}$ is a given  flux expression, prescribed on physical grounds or some other criterion. The symplectic form \eqref{symp C} is not conserved any more and  the symplectic flux at the boundary is given by $\delta\mathbbmss{F}$. Furthermore, with the flux present, the action principle can not be satisfied, 
\begin{align}
    \delta (S+S_b)\approx \int_{\mathscr{I}} \mathbbmss{F}.
\end{align}

On the boundary of AdS spacetime, one can make the flux term $\mathbbmss{F}$ vanishes by imposing boundary conditions. However, as we show in the next section, in presence of large gauge transformations, the flux is generically non-vanishing unless further dynamical constraints are imposed.


\subsection{Renormalization}
In this section, we consider renormalization of the action of a $p$-form theory.  
This is done by introducing an appropriate counter-term  which renders both the on-shell action and  the variation of the action finite. The key point which is pointed out by Comp\'ere and Marolf in \cite{Compere:2008us} is that variation of the counterterm produces divergent corner terms  which also renormalize the symplectic form of the theory. This follows from the Harlow-Wu construction of the symplectic form \eqref{symp C}  for some divergent $C^\mu$ determined from the counter-terms, and leads to a finite symplectic form.

The appropriate counter-term action would be,  
\begin{equation}\label{generalcounterterm}
S_{ct}[R]=\int_{r=R}L_{ct}=\left\{
\begin{array}{l}
    \frac{1}{2(p+1)!}\int_{r=R} \sqrt{g}\A^{[\alpha]}_{\un{\mu}^{p+1}}\F_{[\alpha]}^{r\un{\mu}^{p+1}} \qquad \qquad n_p\geq 2\\\\
    \frac{1}{2(p+1)!}\int_{r=R} \sqrt{g}\A^{[\beta]}_{\un{\mu}^{p+1}}\F_{[\beta]}^{r\un{\mu}^{p+1}} \qquad \qquad n_p< 2
\end{array}\right.
\end{equation}
This in turn leads to $S_\text{ren}=\lim_{R\rightarrow \infty}(S_{reg}[R]+S_{ct}[R])$ and $S_{reg}[R]$ is the bulk action regularized at the cutoff $R$. $S_\text{ren}$ is then the renormalized action.
To fix the renormalization scheme, we assume that the counter term \eqref{generalcounterterm} has no finite term.

The variation of the renormalized action leads to, 
\begin{align}\label{sren vari}
    \delta S_\text{ren}=\int_{\mathscr{I}} \big(\partial_\mu \theta_{ct}^\mu+\Theta\big)
\end{align}
where $\Theta$ is  finite at $R\to\infty$, containing both $A^{[\alpha]}$ and $A^{[\beta]}$ and their variations and it is originated from the bulk action $S$. We dismiss the finite term for the moment and come back to it in the next section. The corner term $\theta^\mu_{ct}$ on the other hand is divergent and is defined by the relation, 
\begin{align}
    \delta L_{ct}=\frac{\delta L_{ct}}{\delta \A_{\un{\mu}^{p+1}}} \delta \A_{\un{\mu}^{p+1}}+\partial_\mu \theta_{ct}^\mu.
\end{align}
In practice the ``equation of motion'' part of $L_{ct}$ cancel the divergent terms in $\theta^r$.  The other divergent term $\partial_\mu \theta_{ct}^\mu$ does not concern the action principle because it is a corner term and is fixed by the initial and final data. However, it leads to a renormalized finite symplectic structure, 
\begin{align}\label{renormalized symplectic form}
    \Omega_\text{ren}=\int_{\Sigma_t} \omega^t_\text{ren}=\int_{\Sigma_t} \omega^t +\oint_{\partial\Sigma_t} \omega^{t}_{ct}.
\end{align}
Explicit calculations for some of the cases are gathered in the appendix \ref{CTAppen}. 


\section{Finite boundary term and different choices of flux}\label{Charge section}
After renormalization of the symplectic form with the corner term $\theta_{ct}$, we turn our attention to the finite term $\Theta$ in \eqref{sren vari} which  turns out to be, 
\begin{align}
\Theta= \frac{c_p\sqrt{h}}{(p+1)!} A_{\un{\nu}^{p+1}}^{[\beta]}\delta A^{\un{\nu}^{p+1}}_{[\alpha]},\qquad\qquad c_p\equiv n_p-1-\delta_{n_p,1}.
\end{align}
Note that after fixing the radial gauge we have the residual gauge transformations, 
$\delta_{\lambda}\A_{\nu\un{\mu}^{p}}=(p+1)\partial_{[\nu}\lambda_{\un{\mu}^{p}]}$ where $\lambda_{\un{\mu}^{p}}$ is independent of $r$. Hence, $A^{[\beta]}$ is gauge invariant while $A^{[\alpha]}\sim\ord{1}$  is not invariant under the residual gauge transformations. 

Using the following decomposition 
\begin{subequations}
\begin{align}
    A_{\mu\un{\nu}^{p}}^{[\alpha]}&=\hat{A}_{\mu\un{\nu}^{p}}^{[\alpha]}+(d\varphi)_{\mu\un{\nu}^{p}}\label{decomp hat alpha}\\
    A_{\mu\un{\nu}^{p}}^{[\beta]}&=\hat{A}_{\mu\un{\nu}^{p}}^{[\beta]}+(d\Phi)_{\mu\un{\nu}^{p}}
\end{align}
\end{subequations}

where $\hat{A}_{\mu\un{\nu}^{p}}^{[\alpha]}$ is gauge invariant and $(d\varphi)_{\mu\un{\nu}^{p}}=(p+1)\partial_{[\mu}\varphi_{\un{\nu}^{p}]}$. We continue to decompose the $\Theta$ term into a gauge invariant term and a term containing $\varphi_{\un{\nu}^{p}}$ which represents the gauge variable of the system. Note that $\Phi_{\un{\nu}^{p}}$ is gauge invariant (see the appendix \ref{Decomposition appendix} for details). The decomposition \eqref{decomp hat alpha} leads to a decomposition in $\Theta$ as We have
\begin{align}
    \Theta =\hat\Theta+\Theta_{\varphi}
\end{align}
where
\begin{align}
      &\hat\Theta=\frac{c_p\sqrt{h}}{(p+1)!} A^{\un{\nu}^{p+1}}_{[\beta]}\delta \hat{A}_{\un{\nu}^{p+1}}^{[\alpha]}\\
      &\Theta_{\varphi}=\frac{c_p\sqrt{h}}{p!} A^{\mu\un{\nu}^{p}}_{[\beta]} D_{[\mu}\delta\varphi_{\un{\nu}^{p}]}.
\end{align}
Now fixing the gauge variable $\varphi$ on the boundary, or assuming that the gauge transformations vanish near the boundary, leads to vanishing of $\Theta_{\varphi}$ term. Therefore, we are left with a gauge-invariant boundary term $\hat{\Theta}$  to be dealt with by imposing boundary conditions . In particular,  by adding finite boundary actions one can eliminate the contribution form $\hat{\Theta}$ and make the action principle well-defined (reviewed in the appendix \ref{Theta gi appendix}). In presence of non-vanishing gauge parameter on the boundary, we are still left with $\Theta_{\varphi}$. Using the language of \eqref{BC-HW-F}, suppose that there is a prescribed flux term $\mathbbmss{F}$ such that 
\begin{align}\label{Varphi Dcomposition}
    \Theta_{\varphi}=\partial_\mu \C^\mu+\mathbbmss{F}
\end{align}
for some $\C^\mu$. In what follows, we examine specific expressions for flux terms $\mathbbmss{F}$ and corresponding surface charges.

\subsection{The Conservative approach $\mathbbmss{F}=0$ }
Among the choices we have, there is a special case where we take the system to be completely conservative. This can be done by taking the decomposition \eqref{Varphi Dcomposition} such as, 
\begin{align}
    \C_{_\circ}^\mu &=\frac{c_p\sqrt{h}}{p!} A^{\mu\un{\nu}^{p}}_{[\beta]}\delta\varphi_{\un{\nu}^{p}}\\
    \mathbbmss{F}_{_\circ} &=0
\end{align}
where we have used the leading term of constraint equation \eqref{Ceq}, $D_\mu A^{\mu\un{\nu}^{p}}_{[\beta]} \approx 0$.  
Then, the symplectic form will be deformed as, 
\begin{align}\label{trivial symplectic form}
    \Omega_{_\circ}=\int_{\Sigma_t} \omega^t_\text{ren}+\oint_{\partial\Sigma_t} \delta \C_{_\circ}^t.
\end{align}
This choice is interesting as guarantees conservation by setting the flux to zero. As it turns out, the surface charges are vanishing
\begin{align}
    \Omega_{_\circ}(\cdot, \delta_\lambda)=\delta Q_\lambda=0.
\end{align}
The reason is that, the corner-term exactly cancels a similar term coming from the bulk symplectic form $\omega^t_\text{ren}$. To show this, using the decomposition \eqref{decomp hat alpha}, dependence of the bulk symplectic form on $\varphi$ can be isolated:
\begin{align}\label{symplectic varphito b}
   \int \omega^t_\text{ren}&=\frac{1}{(p+1)!}\int \sqrt{g}    \delta\A_{\un{i}^{p+1}}\delta \F^{t\un{i}^{p+1}}\nonumber\\
    &=\frac{1}{p!}\int \sqrt{g}    \partial_j\delta \varphi_{\un{i}^{p}}\delta \F^{tj\un{i}^{p}}+\int_{\Sigma_t} \omega^t_{\text{ren,gi}}\nonumber\\
    &\approx \frac{1}{p!}\oint \sqrt{g}   \delta \varphi_{\un{i}^{p}}\delta \F^{tr\un{i}^{p}}+\int_{\Sigma_t} \omega^t_{\text{ren,gi}}\nonumber\\
    &=\frac{c_p}{p!}\oint  \sqrt{q} \delta \varphi_{\un{i}^{p}}\delta A^{t\un{\nu}^{p}}_{[\beta]}+\int_{\Sigma_t} \omega^t_{\text{ren,gi}}
\end{align}
where $\omega^t_{\text{ren,gi}}$ is the renormalized symplectic form free of the gauge variable $\varphi$. In consequence, the total symplectic form is gauge invariant (i.e. annihilated by all gauge transformations)
\begin{align}\label{trivial symplectic form2}
    \Omega_{_\circ}=\int_{\Sigma_t} \omega^t_{\text{ren,gi}}.
\end{align}
The action principle is satisfied  with this choice.

\subsection{Leaky boundary condition $\Delta$}
Another possibility in defining the charges is to allow non-zero flux as 
\begin{subequations}
\begin{align}
    \C_{_\Delta}^\mu &=0\\
    \mathbbmss{F}_{_\Delta} &=\frac{c_p\sqrt{h}}{p!} A^{\mu\un{\nu}^{p}}_{[\beta]} D_{[\mu}\delta\varphi_{\un{\nu}^{p}]}
\end{align}
\end{subequations}
together with the boundary gauge condition, 
\begin{align}\label{boundary gauge fixing}
    \dot{\varphi}_{\un{i}^{p}}=0
\end{align}
where the notation $\Delta$ will be clarified in a moment. With this choice where $\C$ is taken to be zero, the original (renormalized) symplectic form is not deformed,
\begin{align}
    \Omega_{_\Delta} =\int_{\Sigma_t} \omega^t_\text{ren}
\end{align}
and from \eqref{symplectic varphito b} it can easily be seen that the charges are, 
\begin{align}\label{Multipole charges}
    Q_\lambda=\frac{1}{p!} \oint \lambda_{\un{i}^{p}}  F^{rt\un{i}^{p}}_{[\beta]}
\end{align}
where $F^{rt\un{i}^{p}}_{[\beta]}$ is the radial component of the electric field at order $\ord{r^{-n_p}}$. Note that we brought up the indices of $F^{rt\un{i}^{p}}_{[\beta]}$ using the boundary metric $h_{\mu\nu}$ where $F_{rt\un{i}^{p}}^{[\beta]}=-c_p A_{t\un{i}^{p}}^{[\beta]}$. 
Also, recall from section \ref{Wald} that the radial electric field is determined exclusively by the $p$-form components $\phi$. We may cast the expression \eqref{Multipole charges} into another form recalling from \eqref{tr phi} that $F^{[\beta]}_{tr\un{i}^p} \approx \sum \kk_p^2{\phi}^{(k,0)}\mathbb{T}^{(k)}_{\un{i}^{p}}$ where ${\phi}^{(k,0)}$ is the leading $\ord{1}$ order term of $\phi$. Taking the gauge parameter as one of the spherical harmonic $p$-forms $\mathbb{T}^{(k)}_{\un{i}^{p}}$ gives us the charge,  
\begin{equation}\label{charge multipole}
    Q_{(k)}={\kk_p^2} {\phi}^{(k,0)}.
\end{equation}

\paragraph{Flux}  The boundary gauge fixing \eqref{boundary gauge fixing} makes the large gauge transformations time-independent.  Using the constraint equation \eqref{Ceq} we have  $\partial_t A^{t\un{i}^{p}}_{[\beta]}=\Delta \Phi^{\un{i}^{p}}$. In consequence,  we can rewrite the flux term as
\begin{align}
    \mathbbmss{F}_{_\Delta} &=\frac{-c_p\sqrt{h}}{p!} \Delta \Phi^{\un{i}^{p}}\delta\varphi_{\un{i}^{p}}\,.
\end{align}
Therefore, the flux term in proportional to the term $\Delta \Phi^{\un{i}^{p}}$ which justifies the notation $\Delta$. In terms of $\phi$ this is simply $\mathbbmss{F}_{(k)}={\kk_p^2} \dot{{\phi}}^{(k,0)}$. 
Restricting  ourselves to the case where $\dot{\phi}^{(k,0)}=0$ leads to vanishing flux and then  conservation of the multipole charges. This restriction however does not concern the dynamic of the $\psi$ i.e. the $(p+1)$-form sector where their dynamics are decoupled. The interesting situation might be however the electrostatic case, where $\partial_t \F^{tr\un{i}^{p}}=0$ implies that $\Phi^{\un{i}^{p}}=0$ and so vanishing the flux. But this is however, a much smaller phase space on which the above charge are conserved. 

\paragraph{Exact parameters.} The genuine feature of higher form theories with respect to Maxwell theory is the possibility of gauge parameters which are exact; $\laa_{\un{i}^p}=(d\epsilon)_{\un{i}^p}$. Nevertheless, if $F_{[\beta]}^{rt\un{i}^p}$ is a regular $p$-form on sphere, exact parameters will have vanishing charge by equations of motion. 
\paragraph{Harmonic parameters.} The flux term can be written equivalently as,
\begin{align}
    \mathbbmss{F}_{_\Delta} &=\frac{-c_p\sqrt{h}}{p!}  \Phi^{\un{i}^{p}}\delta\Delta\varphi_{\un{i}^{p}}
\end{align}
so it is vanishing for harmonic parameters on the sphere. The gauge parameter $\laa_{\un{\mu}^p}$ can be closed but not exact. We assume the the boundary of $AdS$ has the topology $\mathbb{R}\times K$ where $K$ is a compact manifold. If $K$ is a $n=d-1$ dimensional sphere (as it has been assumed so far), the harmonic forms are either the constant function ($p=0$) or the volume form ($p=n$). The former case is the Maxwell theory in any dimension and the harmonic parameter gives the electric charge in the bulk (the Gauss' law). The second possibility, the field strength $\F$ is a top-form in spacetime. The theory has no propagating degrees of freedom, and the only solution to it is $\F=c\text{Vol}_{d+1}$. The only non-vanishing charge belongs to the harmonic parameter and is proportional to $c$, the value of the constant electric field in space.

There are more interesting possibilities for the harmonic parameter if $K$ has non-trivial topology. If there are $p$-form gauge parameters that are closed but not exact around some non-trivial cycle, the value of the charge gives the electric flux through that cycle.

\subsection{Leaky boundary condition $\Box$}
There is another choice of the flux term which leads to an interesting set of charges. 
Consider the following boundary condition, 
\begin{align}
    &\C_{_{\Box}}^\mu= \frac{c_p\sqrt{h}}{p!}\Big(\hat{A}^{\mu\un{\nu}^{p}}_{[\beta]} \delta\varphi_{\un{\nu}^{p}}+ \Phi_{\un{\nu}^{p}}  (d\delta\varphi)^{\mu\un{\nu}^{p}}\Big)\\
    &\mathbbmss{F}_{_{\Box}}=\frac{-c_p\sqrt{h}}{p!} \partial_t\hat{A}^{t\un{i}^{p}}_{[\beta]} \delta\varphi_{\un{i}^{p}}
\end{align}
where we have used the boundary gauge fixing, 
\begin{align}\label{Bgauge fixing}
    \Box \varphi_{\un{\nu}^{p}} \equiv D^\mu D_{[\mu}\varphi_{\un{\nu}^{p}]}=0,
\end{align}
Now the symplectic form is deformed by the boundary term $\C$ where it deviates from conservation by the above flux term. The symplectic form is given by, 
\begin{align}\label{kappa symplectic form}
    \Omega_{_{\Box}}=\int_{\Sigma_t} \omega^t_\text{ren}+\oint_{\partial\Sigma_t} \delta \C_{_{\Box}}^t.
\end{align}
Using \eqref{symplectic varphito b} and bringing the gauge mode $\varphi$ on the boundary, and working in a temporal gauge-for-gauge condition, $\varphi_{t\un{i}^{p-1}}=0$ we arrive at
\footnote{We can write this symplectic form in terms of the gauge invariant functions $\phi$ and $\psi$ whih would amount to,
\begin{align}\label{Symplectic II phi psi}
\Omega_{_{\Box}}=\int \frac{dr}{(r^2+1)}  \Big({\kk_p^2} \delta \dot{\boldsymbol{\phi}}_{(k)}\delta \boldsymbol{\phi}^{(k)} +\delta \dot{\boldsymbol{\psi}}_{(k)} \delta \boldsymbol{\psi}^{(k)}\Big)+ \Big(\delta \ddot{\phi}^{(k)}  \delta f_{(k)}-\delta \dot{\phi}^{(k)}  \delta \dot{f}_{(k)}\Big)|_{\mathscr{I}}\,.
\end{align}}.
\begin{align}
\Omega_{_{\Box}}=\int_{\Sigma_t} \omega^t_{\text{ren,gi}}+
     \frac{c_p}{(p!)}\oint_{\partial\Sigma_t} \sqrt{q}  \left(\delta\varphi^{\un{i}^p} \delta\dot{\Phi}_{\un{i}^p}+\delta\Phi^{\un{i}^p} \delta\dot{\varphi}_{\un{i}^p}\right)
\end{align}

The charges then will be, 
\begin{align}\label{our charge}
    Q_{\lambda}= \frac{-c_p}{(p!)}\oint_{\partial\Sigma_t} \sqrt{q}  \left(\lambda^{\un{i}^p} \dot{\Phi}_{\un{i}^p}-\Phi^{\un{i}^p} \dot{\lambda}_{\un{i}^p}\right).
\end{align}
In this case we can rewrite the charge in terms of the $\phi$. To do this, note that we had, 
\begin{align}
    F^{[\beta]}_{r\un{i}^{p+1}}=\sum\Big(\dot{{\phi}}^{(k,0)}d\mathbb{T}^{(k)}_{\un{i}^{p}}\Big)
\end{align}
for the $p$-form component (exact component) of the tangent ``magnetic fiel''. In terms of the gauge field, this is just $-c_p (d\Phi)_{\un{i}^{p+1}}$. So we have, 
\begin{align}
    \Phi_{\un{i}^{p}}=\frac{-1}{c_p} \sum\Big(\dot{{\phi}}^{(k,0)}\mathbb{T}^{(k)}_{\un{i}^{p}}\Big)
\end{align}
up to an exact form on the sphere. On the other hand we had the boundary gauge fixing \eqref{Bgauge fixing}. Take $\lambda_{\un{i}^p}=\lambda(t) \mathbb{T}^{(k)}_{\un{i}^{p}}$  subject to 
\begin{align}\label{lambda KK}
    \ddot{\lambda}+\kk_p^2 \laa=0.
\end{align}
The charges \eqref{our charge} may be rewritten as,
\begin{align}\label{our charge1}
    Q_{k}=\lambda\ddot{{\phi}}^{(k,0)}-\dot{\laa}\dot{{\phi}}^{(k,0)}.
\end{align}
Note that for any $k$ we have two independent solution of the equation \eqref{lambda KK} and so two independent charges for any $k$. So the cardinality of this set of charges is twice the previous set \eqref{charge multipole}.  


\paragraph{Flux} Again, using the constraint equations \eqref{Ceq} this time written as $\partial_t \hat{A}^{t\un{i}^{p}}_{[\beta]}+\Box \Phi^{\un{i}^{p}}=0$, we can rewrite the flux as, 
\begin{align}
    \mathbbmss{F}_{_{\Box}}=\frac{c_p\sqrt{h}}{p!} \Box \Phi^{\un{i}^{p}} \delta\varphi_{\un{i}^{p}}.
\end{align}
In term of $\phi$ and using the boundary gauge fixing for $\lambda$, it becomes
\begin{align}
    \mathbbmss{F}_\laa=\lambda\partial_t\left(\ddot{{\phi}}^{(k,0)}+\kk_p^2{\phi}^{(k,0)}\right).
\end{align}
From the equations of motion \eqref{eom1} we have, 
\begin{align}\label{eom phi phi2}
    \ddot{{\phi}}^{(k,0)}+\kk_p^2 {{\phi}}^{(k,0)}=2c_p \phi^{(k,2)}
\end{align}
where $\phi^{(k,2)}$ is the sub-leading component at $r^{-2}$. So we see that the flux of these set of charges is simply, 
\begin{align}
    \mathbbmss{F}_\laa={2c_p }\lambda\dot{\phi}^{(k,2)}
\end{align}
In contrast with the previous set of charges, the requirement of conservation of this charges do not kill the dynamic of $\phi$ completely. We still have the ``boundary dynamic'' of ${\phi}^{(k,0)}$ as is \eqref{eom phi phi2}.       

Cause $\phi$ is the radial electric field, the condition $\dot{\phi}^{(k,2)}=0$ amounts to a constant  radial electric field, \emph{but not at leading order}. Thus, a weaker condition is needed for the charge \eqref{our charge} to be conserved with respect to the standard expression \eqref{Multipole charges}.

\paragraph{Exact charges}
Note that $\varphi_{\un{i}^p}$ and $\Phi_{\un{i}^p}$ is defined up to exact forms on the sphere. So they are in fact in an equivalence class by the relations
\begin{align}
    &\Phi_{\un{i}^p} \sim \Phi_{\un{i}^p}+(d\zeta)_{\un{i}^{p}}\\
        &\varphi_{\un{i}^p} \sim \varphi_{\un{i}^p}+(d\chi)_{\un{i}^{p}}
\end{align}
This ambiguity in $\Phi$ does not change the co-exact charge defined above, because $\lambda$ is defined to be a co-exact form and so the co-exact charge is only  sensitive to the class itself. However there is two set of charges that measure the specific element in the class. Note that after using the exact gauge transformations to set the gauge fixing $\Phi_{t\un{i}^{p-1}}=0$ and $\varphi_{t\un{i}^{p-1}}=0$ we remain with the residual exact gauge transformations,
\begin{align}
    \delta_{\epsilon} \Phi_{\un{i}^p}= (d\epsilon)_{\un{i}^{p}}\\
    \delta_{\tilde{\epsilon}} \varphi_{\un{i}^p}= (d\tilde{\epsilon})_{\un{i}^{p}}
\end{align}
Taking the exact gauge transformations as, $\epsilon_{\un{i}^{p-1}}=\epsilon(t) \mathbb{T}^{(k)}_{\un{i}^{p-1}}$ and $\tilde{\epsilon}_{\un{i}^{p-1}}=\tilde{\epsilon}(t)  \mathbb{T}^{(k)}_{\un{i}^{p-1}}$, the charges will be the followings, 
\begin{align}
    Q^{(k)}_{\epsilon}= c_p \kk_{p-1}^2   \left(\chi^{(k)} \dot{\epsilon}-\epsilon \dot{\chi}^{(k)}\right)\\
     Q^{(k)}_{\tilde{\epsilon}}=c_p  \kk_{p-1}^2  \left(\zeta^{(k)} \dot{\tilde{\epsilon}}-\tilde{\epsilon} \dot{\zeta}^{(k)}\right)
\end{align}
where amounts to the Heisenberg charge algebra, 
\begin{align}
    \{Q^{(k)}_{\epsilon},Q^{(k')}_{\tilde{\epsilon}}\}=c_p \kk_{p-1}^2 \left(\epsilon\dot{\tilde{\epsilon}}-\tilde{\epsilon} \dot{\epsilon}\right) \delta_{kk'}
\end{align}
Note however that these exact charges are not physical because there is no way to measure $\zeta$ and $\chi$. They are just the reflection of the fact that $\Phi$ and $\varphi$ are well-defined up to exact $p$-forms on the sphere. Assuming that the functions $\epsilon(t)$ and $\tilde{\epsilon}(t)$ satisfy the following equation on the boundary, 
\begin{align}
    \ddot{\epsilon}+\kk_{p-1}^2 \epsilon&=0,\\
    \ddot{\tilde{\epsilon}}+\kk_{p-1}^2 \tilde{\epsilon}&=0
\end{align}
leads to conservation of these charges. 

\subsection{A subset of the phase space with conserved charges}
As we saw in the previous section, putting the boundary condition $\dot{\phi}^{(k,2)}=0$ leads to a set of conserved charges in terms of the exact $p$-form components of the tangent magnetic field on the boundary. Now, lets put ${\phi}^{(k,2)}$ and all the sub-leading components of $\phi$ equal to zero. So we only have $\phi={\phi}^{(k,0)}$ which obviously is a asymptotic solution to the equations of motion \eqref{eom1}, if it satisfies,
\begin{align}
   \ddot{{\phi}}^{(k,0)}+\kk_p^2{\phi}^{(k,0)}=0.
\end{align}
These solutions near the boundary can be achieved in terms of the gauge field as well. Consider again the equations of motion \eqref{Deq} in terms of the gauge field in the radial gauge 
\begin{align}\label{Deq Radialgauge1}
     \left[\frac{n_p}{r}+\partial_r\right]\partial_r\A_{\un{\nu}^{p+1}}+r^{-4}\bar{D}^\mu\F_{\mu\un{\nu}^{p+1}}=0
\end{align}
where $\bar{D}^{\mu}=h^{\mu\nu}D_\nu$.   
Consider the following exact solution for the $p+1$-form gauge field for $n_p\neq 1$\footnote{For $n_p=1$, the discussion is similar but the behavior of the gauge field is logarithmic as in \eqref{general falloffs}.}
\begin{align}
    \A_{\mu\un{\nu}^{p}}= r^{1-n_p}(d\Phi)_{\un{\nu}^{p}}
\end{align}
.where $\Phi_{\un{\nu}^{p}}=\Phi_{\un{\nu}^{p}}(x^\mu)$ and we consider $\Phi_{t\un{\nu}^{p-1}}=0$. So, we have $\F_{\mu\un{\nu}^{p+1}}=0$ and this leads to vanishing of the second term of \eqref{Deq Radialgauge1}. The rest of the equations of motion is satisfied with the $r$ profile of the above solution. Note that if we expand this solution in terms of $r$ it only contains the $r^{1-n_p}$ order where the co-exact part is switched off, $\hat{A}_{\mu\un{\nu}^{p}}=0$, and the exact part is just the same $\Phi_{\un{\nu}^{p}}$ here. Note that although $\F_{\mu\un{\nu}^{p+1}}$ is vanishing, the radial electric field and the tangential magnetic field are non-vanishing, 
\begin{align}
   \F_{r\mu \un{\nu}^p} = -c_p r^{-1}\A_{\mu\un{\nu}^{p}}.
\end{align}
There are no subleading terms in this solution, and the dynamics is governed by the constraint equation
\begin{equation}
    \ddot{\Phi}_{\un{i}^p}-\D^k(d\Phi)_{k\un{i}^{p}}=0.
\end{equation}
The surface charge \eqref{our charge} is conserved for this configuration and the integrand is a local function of $\Phi_{\un{i}^p}$.
So they are separately solutions to the first term and the second term of \eqref{Deq Radialgauge}, where the second term is a Maxwell type equations of motion on the boundary. 

Note that we only considered the asymptotic form of the solution here. However, we can interpolate this solution to the bulk and glue that to another form of solution with appropriate sources.

\section{Electric/Magnetic duality and magnetic charges}\label{EM duality section}
The $(p+1)$-form theory with gauge field $A_{\un{I}^{p+1}}$ can be written in terms of the dual gauge field $\tilde{\A}_{\un{I}^{\tilde{p}+1}}$, such that
\begin{equation}\label{duality}
    \tilde{\F}_{\un{I}^{\tilde{p}+2}}=\frac{1}{(p+2)!}{\epsilon^{\un{J}^{p+2}}}_{\un{I}^{\tilde{p}+2}}{\F}_{\un{J}^{p+2}}
\end{equation}
where $\tilde{p}=n-p-2$ and with the same form for the action
\begin{equation}
    S=\frac{1}{2(p+2)!}\int \sqrt{g}{\F}_{\un{J}^{p+2}}{\F}^{\un{J}^{p+2}}=\frac{1}{2(\tilde{p}+2)!}\int \sqrt{g}\tilde{\F}_{\un{J}^{\tilde{p}+2}}\tilde{\F}^{\un{J}^{\tilde{p}+2}}.
\end{equation}

In section \ref{Wald} the $p+1$-form fields were decomposed into spherical harmonics $p$ and $p+1$ forms and the functions $\phi^{(k)}$ and $\psi^{(k)}$ respectively. The normalization \eqref{p normalization} implies that under Hodge duality on $n$-sphere\footnote{See the appendix \ref{Ap-Hodge}.}
\begin{equation}
   \sqrt{\kk_p^2} \,\frac{1}{p!}{\epsilon^{i_1\cdots i_p}}_{j_1\cdots j^{n-p}}\mathbb{T}_{i_1\cdots i_p}^{(k)}=
    (n-p)\partial^{}_{[j_1}\mathbb{T}_{j_2\cdots j^{n-p}]}^{(k)}.
\end{equation}
Using this relation we can read the dual fields as,
\begin{align}\label{phi psi relations duality}
    \tilde{\phi}^{(k)}&=\frac{(-1)^{pd}}{\sqrt{\kk_{p+1}^2}}{\psi}^{(k)},\qquad\qquad\tilde{\psi}^{(k)}=-\sqrt{\kk_p^2}{\phi}^{(k)}.
\end{align}
where $\tilde{\phi}^{(k)}$ and $\tilde{\psi}^{(k)}$ are associated functions of $\tilde{p}$ and $(\tilde{p}+1)$ form components.  

Relations \eqref{phi psi relations duality} are true at any order of $r$ expansion. In terms of the  field strength and the $[\alpha]$ and $[\beta]$ modes we have\footnote{
 $\tilde{A}_{\un{\mu}^{p+1}}^{[\alpha]}$ is not gauge-invariant and only its derivatives appear in duality relations.
}, 
\begin{subequations}
\begin{align}
  \tilde{A}_{\un{\mu}^{p+1}}^{[\beta]} &=\frac{(-1)^{d-p}}{c_p(d-p-1)!}{\epsilon^{\,\un{\nu}^{d-p-1}}}_{\un{\mu}^{p+1}}F^{[\alpha]}_{\un{\nu}^{d-p-1}}\\
  \tilde{F}^{[\alpha]}_{\un{\nu}^{d-p-1}}&= \frac{c_p}{(p+1)!} {\epsilon^{\,\un{\mu}^{p+1}}}_{\,\un{\nu}^{d-p-1}}{A}_{\un{\mu}^{p+1}}^{[\beta]}
\end{align}
\end{subequations}
where the Hodge dual is on the boundary. In addition, under the electric-magnetic duality Dirichlet conditions map to Neumann conditions.

\subsection{Magnetic charges}
The electric charges are  associated with gauge transformations that are non-zero on the boundary. In contrary, it is not possible to obtain magnetic charges from such symmetry principles. However, we can define magnetic charges of a $p+1$-form theory in $AdS_{d+1}$ as follows,
\begin{align}\label{magnetic charge}
    Q^M_\laa=\oint \laa\wedge F^{[\alpha]}=\frac{1}{\tilde{p}!}\frac{1}{(p+2)!}\oint \sqrt{q}\epsilon^{\un{i}^{\tilde{p}}\un{j}^{p+2}} \laa_{\un{i}^{\tilde{p}}} F^{[\alpha]}_{\un{j}^{p+2}}. 
\end{align}
Note that while the gauge parameter of the electric charges (in a $(p+1)$-form gauge theory) is a $p$ form $\laa_{\un{i}^{p}}$,  the parameter in the magnetic charges (in the same theory) is a $\tilde{p}=(n-p-2)$-form $\laa_{\un{i}^{\tilde{p}}}$. Taking  $\laa_{\un{i}^{\tilde{p}}}$ a $p$-form spherical harmonic $\mathbb{T}^{(k)}_{\un{i}^{\tilde{p}}}$ and using the fact that $\kk^2_{\tilde{p}}=\kk^2_{p+1}$ we arrive at, 
\begin{align}\label{Standard M charge}
    Q^M_{(k)}=\sqrt{\kk^2_{p+1}} {\psi}^{(k,0)}
\end{align}
which can be compared to electric charges \eqref{charge multipole}. Note that as mentioned previously, $\psi$ exists only when $n>p+1$, which is equivalent to $\tilde{p}>-1$. If $n=p+1$ (i.e. $\tilde{p}=-1$), the dual theory is a scalar field, thus is has no electric charges. The simplest example is $p=0$, $n=1$, the Maxwell theory in three dimensions.

\paragraph{Magnetic charge as the electric charge of the dual theory} Note that if we reformulate $(p+1)$ form theory as a $(\tilde{p}+1)$ form theory as we did above and seek for the standard electric charge of this dual theory we arrive at, 
\begin{align}
        \tilde{Q}_{(k)}=\frac{1}{p!} \oint \lambda_{\un{i}^{p}}  \tilde{F}^{rt\un{i}^{p}}_{[\beta]}=\kk^2_{\tilde{p}} \tilde{\phi}^{(k,0)}
\end{align}
which is exactly the magnetic charge \eqref{Standard M charge} up to a sign, according to \eqref{phi psi relations duality}. 
So with this observation we advocate the the following definition of the magnetic charges for a $(p+1)$ form theory, 
\begin{align}
    Q^M \equiv \tilde{Q}
\end{align}
where $\tilde{Q}$ are the electric charges of the dual theory which is a $(\tilde{p}+1)$ form theory.
With this proposal we can define other set of magnetic charges as, 
\begin{align}
    Q^M_{(k)}=\frac{1}{\sqrt{\kk^2_{p+1}}}\Big(\lambda \ddot{\psi}^{(k,0)}-\dot{\lambda} \dot{\psi}^{(k,0)}\Big)
\end{align}
where $\lambda$ satisfies, 
\begin{align}
    \ddot{\lambda}+\kk^2_{\tilde{p}}{\lambda}=0.
\end{align}
The flux of this set of charges then will be proportional to $\dot{\psi}^{(k,2)}$. 
In terms of the gauge field, this charges are coming from the co-exact sector of the tangent electric field at the order $\ord{1}$. 

Note that, we can define the magnetic charges as Noether charges of the dual theory, but in the dual description, the notion of electric charges as Noether charges of the original theory is lost. 
We also can go for the dual description only for the $\psi$ sector and have both charges as Noether charges, but we can not maintain the covariance of the theory. For a recent formulation of magnetic charges as the charges of a Noether symmetry, see \cite{Geiller:2021gdk} where they have obtained magnetic charges from first order formulation of $p$-form theory in terms of Abelian BF theory.

\paragraph{Duality invariance in $\boldsymbol{n_p=2}$ }
If $d=2p+3$, the Hodge dual field strength tensors have the same rank as $p=\tilde{p}$. This happens for Maxwell theory in four dimensions ($p=0, d=3$), 2-form theory in six dimensions $(p=1,d=5)$ etc. 
Note also that, under the duality the gauge invariant part of the symplectic form \eqref{sympformphipsi} (i.e. the bulk term) maps to itself. In fact, in these cases the system has a $SO(2)$ symmetry which is a rotation in $\psi$ and $\phi$ variables.
This symmetry is however broken in the presence of the boundary and the gauge mode $\varphi$ ($f$ in \eqref{T decompose}) on it, as the term $(\kk_p^2\delta \phi^{(k)}\delta f_{(k)})|_{\mathscr{I}}$ in \eqref{sympphipsi} or the term $( \delta \ddot{\phi}^{(k)}  \delta f_{(k)}-\delta \dot{\phi}^{(k)}  \delta \dot{f}_{(k)})|_{\mathscr{I}}$ in \eqref{Symplectic II phi psi} are not invariant under this duality. We can however construct symplectic forms that are completely $SO(2)$ invariant by introducing the magnetic gauge variable $g$ which is naturally conjugate to the $\psi$ on the boundary. For example, in the standard form \eqref{sympphipsi} we can add the term, $(\delta \psi^{(k)}\delta g_{(k)})|_{\mathscr{I}}$ which leads to a duality invariant symplectic form for a theory with $n_p=2$, 
\begin{align}
    \Omega_{di}=\int \frac{dr}{(r^2+1)}  \Big({\kk_p^2} \delta \dot{\boldsymbol{\phi}}_{(k)}\delta \boldsymbol{\phi}^{(k)} +\delta \dot{\boldsymbol{\psi}}_{(k)} \delta \boldsymbol{\psi}^{(k)}\Big)+ \Big(\kk_p^2 \delta {\phi}^{(k)}  \delta f_{(k)}+\delta {\psi}^{(k)}  \delta {g}_{(k)}\Big)|_{\mathscr{I}}.
\end{align}
Then, we see that this symplectic form is invariant under $\delta_\theta$ which is the infinitesimal generator of the $SO(2)$ symmetry and is defined as,
\begin{subequations}
\begin{align}
    \delta_\theta \sqrt{\kk_p^2} \boldsymbol{\phi}^{(k)}=\boldsymbol{\psi}^{(k)}\, \qquad
    \delta_\theta \boldsymbol{\psi}^{(k)} =-\sqrt{\kk_p^2} \boldsymbol{\phi}^{(k)}\\
    \delta_\theta \sqrt{\kk_p^2}f^{(k)}=g^{(k)}\, \qquad
    \delta_\theta g^{(k)} =- \sqrt{\kk_p^2}f^{(k)}
\end{align}
\end{subequations}
where the charge will be, 
\begin{align}
    Q_\theta=\sqrt{\kk_p^2} \int \frac{dr}{(r^2+1)}  \Big( \dot{\boldsymbol{\phi}}_{(k)} \boldsymbol{\psi}^{(k)} -\dot{\boldsymbol{\psi}}_{(k)} \boldsymbol{\phi}^{(k)}\Big)+ \sqrt{\kk_p^2}\Big( {\phi}^{(k)}  g_{(k)}- {\psi}^{(k)}   f_{(k)}\Big)|_{\mathscr{I}} 
\end{align}
with the algebra, 
\begin{align}
    \{Q_\theta,Q\}=Q^M,\qquad\qquad \{Q_\theta,Q^M\}=-Q.
\end{align}

\section{Integrability}\label{energy}
The electric surface charges in $p+1$-form theory are integrable as long as the gauge parameter is not field-dependent. This is not the case for the charges of AdS isometries. For instnace, including boundary modes modifies the energy of the system by surface terms.  As it turns out, the general form of the energy computed  as $\Omega(\cdot, \delta_t)$ will not be integrable in the presence of the gauge mode $f$ (or $\varphi$) by a non-integrable boundary term. The integrable bulk term is universal and is given by the standard energy-momentum tensor as expected,  
\begin{align}
    E_{\text{bulk}}= \frac12 \int \frac{dr}{(r^2+1)}  \Bigl[{\kk_p^2}\Big(\dot{\boldsymbol{\phi}}^2 +\boldsymbol{\phi} A_{0}\boldsymbol{\phi}\Big) + \Big(\dot{\boldsymbol{\psi}}^2 +\boldsymbol{\psi} A_{1}\boldsymbol{\psi}\Big)\Bigr] 
\end{align}
where $A_{0}$ and $A_{1}$ are differential operators in the equations of motion \eqref{phieomm} and \eqref{psieqx} as $\ddot{\boldsymbol{\phi}}=-A_{0}\boldsymbol{\phi}$ and $\ddot{\boldsymbol{\psi}}=-A_{1}\boldsymbol{\psi}$.  Sum over the different spherical modes $(k)$ is  assumed here. The extra non-integrable part denoted as $\slashed{\delta}E$ is different using different symplectic forms. We have, 
\begin{subequations}
\begin{align}
    \slashed{\delta}E_\circ &=0\\
     \slashed{\delta}E_\Delta &= -\sum\kk_p^2 \dot{\phi}^{(k,0)}\delta f_{(k)}=-\mathbbmss{F}_{\Delta} \\
      \slashed{\delta}E_\Box &= -2c_p\sum\dot{\phi}^{(k,2)}\delta f_{(k)}=-\mathbbmss{F}_{\Box}.
\end{align}
\end{subequations}
Clearly, the energy is integrable when the associated fluxes are vanishing and the surface charges are conserved. 
In the third case, there is also an extra integrable contribution from the boundary modes to the energy,
\begin{align}
    E_\Box=\sum \ddot{\phi}^{(k,0)} \dot{f}_{(k)}+\kk_p^2 \dot{\phi}^{(k,0)} f_{(k)}\,.
\end{align}

\section{Concluding Remarks}
In this paper we computed the  surface charges for a $(p+1)$-form theory on AdS$_{d+1}$ associated with the gauge mode on the boundary of spacetime assuming the usual  reflective boundary condition modulo residual gauge transformations. This results in infinite dimensional Abelian symmetry algebra. Different expressions for the charge are possible depending on the flux expression, leading to different boundary terms of the symplectic form. We studied three cases of interest where the results summarized in table \ref{Summary table} . The first one was a conservative boundary condition which resulted in an empty set  of charges. Similarly to the recent gravity studies \cite{Compere:2019bua}, here we see that demanding infinite dimensional asymptotic charges in incompatible with conservative boundary conditions. So, in two of these cases, we choose a leaky boundary condition where there is a non-trivial flux through the boundary unless further assumptions on the dynamics is made.

By proposing  a generic counter-term, we showed how the divergences in the on-shell action are subtracted in various cases. In addition, we explicitly showed how variation of the counter term on the boundary leads to  divergent corner terms  which render the symplectic form finite. This procedure has also been applied in Maxwell theory in flat space \cite{Freidel:2019ohg}. The charges are then finite by construction as they are built out of the renormalized symplectic form.

\begin{table}[h]
\centering
    \begin{tabular}{|c||c|c|c|}
    \hline
     &$\circ$& $\Delta$&$\Box$\\    \hline
    $\mathbbmss{F}$ &   $0$     &  $ \Delta \Phi^{\un{i}^{p}}\delta\varphi_{\un{i}^{p}}$ &  $ \Box \Phi^{\un{i}^{p}}\delta\varphi_{\un{i}^{p}}$  \\    \hline
     $\Omega$   & $\int \omega^t_{\text{ren,gi}}$ & $\int \omega^t_\text{ren}$& $ \int \omega^t_{\text{ren,gi}}+\frac{c_p}{(p!)}\oint \sqrt{q}  \left(\delta\varphi^{\un{i}^p} \delta\dot{\Phi}_{\un{i}^p}+\delta\Phi^{\un{i}^p} \delta\dot{\varphi}_{\un{i}^p}\right)$ \\    \hline
    $Q_{(k)}$  & $\emptyset$ & ${\kk_p^2} {\phi}^{(k,0)}$ & $\Big(\lambda\ddot{{\phi}}^{(k,0)}-\dot{\laa}\dot{{\phi}}^{(k,0)}\Big)$ \\   \hline 
    \end{tabular}
    \caption{This table summarizes our main results. $\mathbbmss{F}$ is the flux. $\Phi^{\un{i}^{p}}$ is the exact part of the tangent ``magnetic field'' $\F_{r\un{i}^{p+1}}$.  $\varphi_{\un{i}^{p}}$ is the gauge mode and ${\phi}^{(k,0)}$ is the gauge invariant $p$ component defined in \eqref{F_ai to phi}. $\Delta$ and $\Box$ are also the Laplacian operators  respectively on the sphere and on the cylinder. $Q_{(k)}$ is the surface charge smeared on  $p$-form spherical harmonics.}   
    \label{Summary table}
\end{table}

Boundary conditions advocated here are similar to the usual reflective boundary conditions, as the gauge field on the boundary is fixed up to gauge transformations. Energy as defined by the bulk integration of the energy-momentum tensor is finite and conserved. However, covariant phase space formalism predicts a boundary term for the energy depending on the pure gauge mode. This new piece spoils integrability and conservation of the energy despite  the lack of  radiation through the boundary.

Finally, we only considered Maxwell-like actions, whereas Chern-Simons theories can be studied along the same lines. These theories besides being  physically interesting,  have a rich asymptotic structure and boundary dynamics.

\section*{Acknowledgements}
We are especially grateful to Shahin Sheikh-Jabbari for collaboration in early stages of this work and also for careful reading of the manuscript. We are thankful to  Romain Ruzziconi for comments on the draft and to Reza Javadinezhad, Ali Naseh, Ali Seraj, Hesam Soltanpanahi and Behrad Taghavi for helpful discussions. EE was supported in part by  Saramadan grant number ISEF/M/99376.

\appendix

\section{Holographic renormalization of action and the symplectic form}\label{CTAppen}
We show that adding the counter term \eqref{generalcounterterm} makes the on-shell action, variation of the action, and the symplectic form finite.
\subsection*{On-shell action and variation of the action}
\begin{equation}
    S_{\text{reg}}[R]\approx-
    \frac{1}{2(p+1)!}\int_{r=R} d^{d}x \sqrt{g}\A_{\un{\mu}^{p+1}}\F^{r\un{\mu}^{p+1}}
\end{equation}
According to fall-offs \eqref{general falloffs}, the divergent terms come from $A_{[\alpha]}$ modes in $n_p\geq 2$ and from $A_{[\beta]}$ modes in $n_p<2$. Thus, the counterterm \eqref{generalcounterterm} cancels the divergences trivially. This leads to, 
\begin{align}
S^{\text{on-shell}}_{\text{ren}}=\frac{c_p}{2(p+1)!} \int_{\mathscr{J}} d^{d}x \sqrt{h}A_{\un{\nu}^{p+1}}^{[\beta]} A^{\un{\nu}^{p+1}}_{[\alpha]}
\end{align}
On the other hand, the variation of the renormalized action is turned out to be, 
\begin{align}
\delta S_{\text{ren}}=\frac{c_p}{(p+1)!} \int_{\mathscr{J}} d^{d}x \sqrt{h}A_{\un{\nu}^{p+1}}^{[\beta]} \delta A^{\un{\nu}^{p+1}}_{[\alpha]}
\end{align}

\subsection*{Symplectic form}
In the radial gauge, the symplectic form is, 
\begin{align}
    \Omega=\frac{1}{(p+1)!}\int \sqrt{g} d^nx^i dr   \delta\A_{\un{i}^{p+1}}\delta \F^{t\un{i}^{p+1}} =\frac{1}{(p+1)!}\int \sqrt{q} d^nx^i dr \frac{r^{n_p-2}}{r^2+1} \delta \F_{t\un{i}^{p+1}} \delta\A_{\un{i}^{p+1}}
\end{align}
where the term $\delta\A_{r\un{i}^{p}}\delta \F^{tr\un{i}^{p}}$ drops out of the symplectic form. This has crucial consequences. First note that in this gauge fixed approach, the $p$-form and $(p+1)$-form components have the same  contribution in the bulk symplectic form. One can see the symplectic form as a $p$-form sector plus a $(p+1)$-from sector, $\delta\A^{(p)}_{\un{i}^{p+1}}\delta \F_{(p)}^{t\un{i}^{p+1}}+\delta\A^{(p+1)}_{\un{i}^{p+1}}\delta \F_{(p+1)}^{t\un{i}^{p+1}}$ where the term $\delta\A^{(p)}_{r\un{i}^{p}}\delta \F_{(p)}^{tr\un{i}^{p}}$ is absent due to radial gauge. We will see that for example in $n_p=0, 1$ this fact leads to the conclusion that we have two normalizable modes for both $p$ and $(p+1)$-form sector, in contrast to the gauge invariant approach where we were working with gauge invariant quantities $\psi$ and $\phi$ in the section \ref{NORSection}.  Specifically in the table \ref{NNMs}, we had $2$ normalizable modes for $p+1$ components and $1$ normalizable mode for $p$-form components sector in that gauge invariant approach.

Now we review how the holographic renormalization of the action eliminates some divergences of the symplectic form and leads to finite one in some cases.

\paragraph{$\boldsymbol{n_p=0}$} 
While the action is divergent in this case and need to be renormalized, the symplectic form is finite as we explained above. One can see this by writing the symplectic form expansion, 
\begin{align}
    \Omega=\int \sqrt{q} d^nx^i dr r^{-4}(1-r^{-2}+\cdots) \delta (F^{[\beta]}_{t\un{i}^{p+1}} r+ \cdots) \delta(A_{[\beta]}^{\un{i}^{p+1}} r+ \cdots) \sim \text{finite} 
\end{align}
where the leading term in the symplectic density is $\ord{r^{-2}}$ and so normalizable. On the other hand, the variation of the action has the following divergent term, 
\begin{align}
    \delta S_{div} \approx \frac{-1}{(p+1)!}\int_{\mathscr{I}}\sqrt{h}\left(r  A^{[\beta]}_{\un{\nu}^{p+1}} \delta  A_{[\beta]}^{\un{\nu}^{p+1}}\right)
\end{align}
which is cured by the renormalization. 

\paragraph{$\boldsymbol{n_p=1}$} 
This case is also like the previous one. One can see that the symplectic form is convergent thanks to the radial gauge, 
\begin{align}
  \Omega=\int \sqrt{q} d^nx^i dr r^{-3}(1-r^{-2}+\cdots) \delta (F^{[\beta]}_{t\un{i}^{p+1}} \ln{r}+ \cdots) \delta(A_{[\beta]}^{\un{i}^{p+1}} \ln{r}+ \cdots) \sim\text{finite}  
\end{align}
where the leading term in the symplectic density is $\ord{r^{-3}\ln^2 {r}}$. The divergent term of the variation of the action is,
\begin{align}
    \delta S_{div} \approx \frac{-1}{(p+1)!}\int_{\mathscr{I}}\sqrt{h}\left(\ln{r}  A^{[\beta]}_{\un{\nu}^{p+1}} \delta  A_{[\beta]}^{\un{\nu}^{p+1}}\right)
\end{align}
which is again cured by the renormalization. 

\paragraph{$\boldsymbol{n_p=2}$}
The case $n_p=2$ is a special case. This case has had already two independent normalizable modes in terms of $\psi$ and $\phi$ and the action is also finite and does not need renormalization. The symplectic form is also finite, 
\begin{align}
  \Omega=\int \sqrt{q} d^nx^i dr r^{-2}(1-r^{-2}+\cdots) \delta (F^{[\alpha]}_{t\un{i}^{p+1}}+ \cdots) \delta(A_{[\alpha]}^{\un{i}^{p+1}}+ \cdots) \sim\text{finite}  
\end{align}
where the leading term in the symplectic density is $\ord{r^{-2}}$.

\paragraph{$\boldsymbol{n_p=3}$}
In this case the symplectic form and also the action diverge and it needs a renormalization. 
We explain in detail the renormalization of this case in detail which almost contains all the possible situations in a simple way. Considering the expansion of the gauge field as, 
\begin{align}
    \A=(A^{[\alpha]}+ A^{[\alpha,2]}r^{-2}\ln{r}+\cdots)+(A^{[\beta]} r^{-2}+ \cdots)
\end{align}
where we suppressed the indices. Then from the equations of motion \eqref{Deq}, in radial gauge we have, 
\begin{align}\label{Deq Radialgauge}
     \left[\frac{n_p}{r}+\partial_r\right]\partial_r\A_{\un{\nu}^{p+1}}+r^{-4}\bar{D}^\mu\F_{\mu\un{\nu}^{p+1}}=0
\end{align}
where $\bar{D}^{\mu}=h^{\mu\nu}D_\nu$. This leads to, 
\begin{align}\label{EQM02}
    A^{[\alpha,2]}_{\un{\mu}^{p+1}}=\frac{1}{2}D_\mu F^{\mu\un{\mu}^{p+1}}_{[\alpha]}
\qquad\qquad
    F_{\mu\un{\mu}^{p+1}}^{[\alpha]}\equiv(p+2) D_{[\mu} A^{[\alpha]}_{\un{\mu}^{p+1}]}.
\end{align}
Now the variation of the action reads, 
\begin{align}
    \delta S \approx \frac{-1}{(p+1)!}\int_{\mathscr{I}}\sqrt{h}\left( (1-2\ln{r}) A^{[\alpha,2]}_{\un{\nu}^{p+1}}\delta A_{[\alpha]}^{\un{\nu}^{p+1}}+2 A_{[\beta]}^{\un{\nu}^{p+1}}\delta A^{[\alpha]}_{\un{\nu}^{p+1}}\right)
\end{align}
which has the logarithmic divergence in the first term. As proposed in \eqref{generalcounterterm} the suitable counter-term action is, 
\begin{align}
    S_{ct}=\frac{1}{2(p+1)!}\int d^{d}x^\mu \sqrt{g}\A^{[\alpha]}_{\un{\mu}^{p+1}}\F_{[\alpha]}^{r\un{\mu}^{p+1}}
\end{align}
which leads to, 
\begin{align}
    \delta S_\text{ren} \approx \frac{1}{(p+1)!}\int_{\mathscr{I}}\sqrt{h}\left\{\frac{(2\ln{r}-1)}{2}\left( A^{[\alpha,2]}_{\un{\nu}^{p+1}}\delta A_{[\alpha]}^{\un{\nu}^{p+1}}-A^{[\alpha]}_{\un{\nu}^{p+1}}\delta A_{[\alpha,2]}^{\un{\nu}^{p+1}}\right)+2 A_{[\beta]}^{\un{\nu}^{p+1}}\delta A^{[\alpha]}_{\un{\nu}^{p+1}}\right\}
\end{align}
Our goal here is to use the equation of motion \eqref{EQM02} to rewrite the variation of the renormalized action as, 
\begin{align}
    \delta S_\text{ren} \approx \int_{\mathscr{I}} \partial_\mu \theta_{ct}^\mu + \Theta
\end{align}
where $\Theta$ is a finite term and all possible divergent terms are in $\theta_{ct}$ which is a corner term. 
It is straightforward that using the equation of motion,
\begin{align}
    \delta S_\text{ren} \approx \frac{1}{(p+1)!}\int_{\mathscr{I}}\sqrt{h}\left\{\frac{(2\ln{r}-1)}{4}D_\mu\left(  F^{\mu\un{\nu}^{p+1}}_{[\alpha]}\delta A^{[\alpha]}_{\un{\nu}^{p+1}}-A^{[\alpha]}_{\un{\nu}^{p+1}}\delta F^{\mu\un{\nu}^{p+1}}_{[\alpha]}\right)+2 A_{[\beta]}^{\un{\nu}^{p+1}}\delta A^{[\alpha]}_{\un{\nu}^{p+1}}\right\}
\end{align}
Where we can identify, 
\begin{align}
   &\theta^\mu_{ct}= \frac{1}{(p+1)!}\sqrt{h}\frac{(2\ln{r}-1)}{4}\left(  F^{\mu\un{\nu}^{p+1}}_{[\alpha]}\delta A^{[\alpha]}_{\un{\nu}^{p+1}}-A^{[\alpha]}_{\un{\nu}^{p+1}}\delta F^{\mu\un{\nu}^{p+1}}_{[\alpha]}\right)\\
   &\Theta=\frac{1}{(p+1)!}\sqrt{h} 2 A_{[\beta]}^{\un{\nu}^{p+1}}\delta A^{[\alpha]}_{\un{\nu}^{p+1}}
\end{align}

Now the symplectic form has the following expansion,  
\begin{align}
  \Omega=\int \sqrt{q} d^nx^i dr r^{-1}\delta F^{[\alpha]}_{t\un{i}^{p+1}}\delta A_{[\alpha]}^{\un{i}^{p+1}}+ \text{finite}
\end{align}
where the leading term has logarithmic divergence. 
Then, it is then easily seen that the counter term symplectic potential renormalizes the symplectic form by the relation, 
\begin{align}
    \Omega_\text{ren}=\int_{\Sigma_t} \omega^t +\oint_{\partial\Sigma_t} \omega^{t}_{ct}
\end{align}
where 
\begin{align}
    \omega^{\mu}_{ct}=\frac{1}{(p+1)!}\sqrt{h}\ln{r} \delta F^{\mu\un{\nu}^{p+1}}_{[\alpha]}\delta A^{[\alpha]}_{\un{\nu}^{p+1}}
\end{align}

\paragraph{$\boldsymbol{n_p=4}$}
This case is almost similar to the previous $n_p=3$. The symplectic form is divergent, 
\begin{align}
  \Omega=\int \sqrt{q} d^nx^i dr \delta F^{[\alpha]}_{t\un{i}^{p+1}}\delta A_{[\alpha]}^{\un{i}^{p+1}}+ \text{finite}  
\end{align}
With renormalization one lead to,
\begin{align}
   &\theta^\mu_{ct}= \frac{1}{(p+1)!}\sqrt{h}\frac{r}{2}\left(  F^{\mu\un{\nu}^{p+1}}_{[\alpha]}\delta A^{[\alpha]}_{\un{\nu}^{p+1}}-A^{[\alpha]}_{\un{\nu}^{p+1}}\delta F^{\mu\un{\nu}^{p+1}}_{[\alpha]}\right)
\end{align}
which leads to a finite symplectic form. 

\paragraph{$\boldsymbol{n_p=5}$}
The same procedure is repeated here. Consider the falloff condition for the gauge field as,
\begin{align}
    \A=(A^{[\alpha]}+A^{[\alpha,2]} r^{-2}+A^{[\alpha,4]} r^{-4}\ln{r}+\cdots)+(A^{[\beta]} r^{-4}+ \cdots)
\end{align}
where from the equations of motion, we have, 
\begin{align}
    A^{[\alpha,2]}_{\un{\mu}^{p+1}}=\frac{1}{4}D_\mu \bar{F}^{\mu\un{\mu}^{p+1}}_{[\alpha]}, \qquad \qquad A^{[\alpha,4]}_{\un{\mu}^{p+1}}=\frac{1}{4}D_\mu \bar{F}^{\mu\un{\mu}^{p+1}}_{[\alpha,2]}-\frac{1}{4}D_\mu \bar{F}^{\mu\un{\mu}^{p+1}}_{[\alpha]}
\end{align} 

Then we have 
\begin{align}
    \delta S_\text{ren} &\approx \frac{1}{(p+1)!}\int_{\mathscr{I}}\sqrt{h}[r^2\left( A^{[\alpha,2]}_{\un{\nu}^{p+1}}\delta A_{[\alpha]}^{\un{\nu}^{p+1}}-A^{[\alpha]}_{\un{\nu}^{p+1}}\delta A_{[\alpha,2]}^{\un{\nu}^{p+1}}\right)\nn\\&+\frac{(1-4\ln{r})}{2}\left( A^{[\alpha]}_{\un{\nu}^{p+1}}\delta A_{[\alpha,4]}^{\un{\nu}^{p+1}}-A^{[\alpha,4]}_{\un{\nu}^{p+1}}\delta A_{[\alpha]}^{\un{\nu}^{p+1}}\right)+4 A_{[\beta]}^{\un{\nu}^{p+1}}\delta A^{[\alpha]}_{\un{\nu}^{p+1}}]
\end{align}

where the symplectic form is, 
\begin{align}
  \Omega=\int \sqrt{q} d^nx^i dr\left\{ r \delta F^{[\alpha]}_{t\un{i}^{p+1}}\delta A_{[\alpha]}^{\un{i}^{p+1}}+ r^{-1}\left(\delta F^{[\alpha,2]}_{t\un{i}^{p+1}}\delta A_{[\alpha]}^{\un{i}^{p+1}}+\delta F^{[\alpha]}_{t\un{i}^{p+1}}\delta A_{[\alpha,2]}^{\un{i}^{p+1}}-\delta F^{[\alpha]}_{t\un{i}^{p+1}}\delta A_{[\alpha]}^{\un{i}^{p+1}}\right)\right\}+\text{finite}  
\end{align}

and using them we lead to, 

\begin{align}
         \theta_{ct}^\mu&= \frac{\sqrt{h}}{(p+1)!}\Bigl\{\frac{r^2}{4}(A^{[\alpha]}_{\un{\nu}^{p+1}}\delta F^{\mu\un{\nu}^{p+1}}_{[\alpha]}-F^{\mu\un{\nu}^{p+1}}_{[\alpha]}\delta A^{[\alpha]}_{\un{\nu}^{p+1}}) \nn \\
         &+\frac{(4\ln{r}-1)}{8}\Bigl[(F^{\mu\un{\nu}^{p+1}}_{[\alpha,2]}\delta A^{[\alpha]}_{\un{\nu}^{p+1}}-A^{[\alpha]}_{\un{\nu}^{p+1}}\delta F^{\mu\un{\nu}^{p+1}}_{[\alpha,2]})+(F^{\mu\un{\nu}^{p+1}}_{[\alpha]}\delta A^{[\alpha,2]}_{\un{\nu}^{p+1}}-A^{[\alpha,2]}_{\un{\nu}^{p+1}}\delta F^{\mu\un{\nu}^{p+1}}_{[\alpha]})\nn\\&-(F^{\mu\un{\nu}^{p+1}}
         _{[\alpha]}\delta A^{[\alpha]}_{\un{\nu}^{p+1}}-A^{[\alpha]}_{\un{\nu}^{p+1}}\delta F^{\mu\un{\nu}^{p+1}}_{[\alpha]})\Bigr]\Bigr\}
\end{align}
This counter term cancels the divergent terms in the symplectic form, leading to a renormalized symplectic form.

\paragraph{$\boldsymbol{n_p>5}$ and $\boldsymbol{n_p<0}$}
These cases can also be renormalized using the same method.

\section{Decomposition of the gauge field} \label{Decomposition appendix} 
Note that we have decomposed the gauge field in terms of the $p$-form and $(p+1)$-form components on the sphere. For the spherical components we then write, 
\begin{align}
    A^{[\alpha]}_{\un{i}^{p+1}}=\hat{A}^{[\alpha]}_{\un{i}^{p+1}}+(p+1)D_{[i}\varphi_{\un{i}^{p}]}
\end{align}
Where $\hat{A}$ is the co-exact $p+1$-form component which is gauge invariant. Then we define, $\hat{A}^{[\alpha]}_{t\un{i}^{p}}=A^{[\alpha]}_{t\un{i}^{p}}-\partial_t \varphi_{\un{i}^{p}}$ which with previous decomposition we sum up as, 
\begin{align}\label{DECom alpha}
    A_{\mu\un{\nu}^{p}}^{[\alpha]}=\hat{A}_{\mu\un{\nu}^{p}}^{[\alpha]}+(p+1)D_{[\mu}\varphi_{\un{\nu}^{p}]}
\end{align}
where the temporal gauge fixing $\varphi_{t\un{i}^{p-1}}=0,$ must be understood here. This gauge fixing is accessible by exact gauge transformation $\delta_{\epsilon}\varphi_{\nu\un{\mu}^{p-1}}=p\partial_{[\nu}\epsilon_{\un{\mu}^{p-1}]}$.
Then we conclude that $\hat{A}^{[\alpha]}_{t\un{i}^{p}}$ is also gauge invariant. So the decomposition \eqref{DECom alpha} can also be seen as decomposition to gauge invariant and pure gauge part. 
We can do the same decomposition for $A_{\nu\un{\mu}^{p}}^{[\beta]}$. We write, 
\begin{align}\label{DECom beta}
    A_{\mu\un{\nu}^{p}}^{[\beta]}=\hat{A}_{\mu\un{\nu}^{p}}^{[\beta]}+(p+1)D_{[\mu}\Phi_{\un{\nu}^{p}]}
\end{align}
where again we have $\Phi_{t\un{i}^{p-1}}=0$. However note that in this case $A^{[\beta]}$ is gauge invariant and then so $\Phi$ and $\hat{A}^{[\beta]}$. The constraint equations are $D_\mu A_{[\beta]}^{\mu\un{\nu}^{p}}=0$ so that
\begin{align}\label{Constraint beta}
    D_\mu A_{[\beta]}^{\mu\un{i}^{p}}=\partial_t\hat{A}^{t\un{i}^{p}}_{[\beta]}+D_\mu D^{[\mu}\Phi^{\un{i}^{p}]}=0\\
    D_k A_{[\beta]}^{kt\un{i}^{p-1}}=D_k \hat{A}_{[\beta]}^{kt\un{i}^{p-1}}-\partial_tD_k \Phi^{k\un{i}^{p-1}}=0
\end{align}

\section{The gauge independent part of $\Theta$} \label{Theta gi appendix}
Here we examine the flux term $\hat{\Theta}$ . This leads to the symplectic flux for the gauge invariant sector $\hat{\omega} \sim \delta A^{\un{\nu}^{p+1}}_{[\beta]}\delta \hat{A}_{\un{\nu}^{p+1}}^{[\alpha]}$. The fact is that although we have two independent normalizable modes, we must choose one out of these two to have vanishing symplectic flux at least in the absence of $\varphi$ modes which is a pure gauge mode. 
The easy ones are, $\delta A^{[\beta]}=0$ or $\delta \hat{A}^{[\alpha]}=0$. However we have infinite continuous range of choices between these two when when we have two independent normalizable. This is the case for any $p$-form theory after renormalizing the symplectic structure. Here, following \cite{Marolf:2006nd}, if we want $\hat{\omega}$ to be vanished, we assume one of the following boundary conditions, 
\begin{align}
A_{\un{\nu}^{p+1}}^{[\beta]}=J_{\un{\nu}^{p+1}}^{[\beta]}(x, \hat{A}_{\un{\nu}^{p+1}}^{[\alpha]}) \qquad \text{or}\qquad \hat{A}_{\un{\nu}^{p+1}}^{[\alpha]}=J_{\un{\nu}^{p+1}}^{[\alpha]}(x, A_{\un{\nu}^{p+1}}^{[\beta]})
\end{align}
where, $J^{[\beta]}$ and $J^{[\alpha]}$ are ultra-local functions of $\hat{A}^{[\alpha]}$ and $A^{[\beta]}$. 
Vanishing the symplectic flux then indicates the existence of the following potentials in each case, 
\begin{align}
J^{\un{\nu}^{p+1}}_{[\beta]}=
     \frac{(p+1)!}{c_p\sqrt{h}}\frac{\delta W_{[\beta]}}{\delta \hat{A}_{\un{\nu}^{p+1}}^{[\alpha]}}, \qquad \qquad J^{\un{\nu}^{p+1}}_{[\alpha]}=
     \frac{(p+1)!}{c_p\sqrt{h}}\frac{\delta W_{[\alpha]}}{\delta {A}_{\un{\nu}^{p+1}}^{[\beta]}}
\end{align}
Introducing these functions $W$ is to ensure that we have a well-defined action principle at least for this gauge invariant sector. We go on by introducing, 
\begin{align}
    S_W=\int_{\mathscr{I}} W
\end{align}
Then on can show that,
\begin{align}
    \delta(S_\text{ren}+S_W)=\int \Theta_{\varphi}
\end{align}
So whenever we ignore the boundary gauge mode, by putting $\delta \varphi=0$ leading to $\Theta_{\varphi}=0$, we have a well-defined action principle.

\section{Spherical Harmonics $p$-forms under Hodge duality}\label{Ap-Hodge}
Let us start by some basic definitions. Consider the inner product of $p$-forms on a compact $n$-manifold
\eqs{
\langle\alpha,\beta\rangle=\int_\M\alpha\wedge\star\beta\,.\label{inner product}
}
The adjoint of the exterior derivative $\text{d}$ with respect to this inner product is called the co-differential operator $\di^\dagger$, related to the exterior derivative by
\eqs{\label{codif}
\di^\dagger=(-1)^{n(p+1)+1}\star\di\star\,.}
 The Laplace-Beltrami operator $\Delta$ is a second-order differential operator, mapping $p$-forms to $p$-forms
\eqs{
\Delta=\di\di^\dagger+\di^\dagger\di\,.
}
A differential form $\alpha$ is called harmonic if $\Delta\alpha=\di\alpha=\di^\dagger\alpha=0\,.$ According to the Hodge decomposition theorem, any differential $p$-form  on a closed Riemannian manifold  can be decomposed into exact, co-exact and harmonic forms:
\eqs{
\omega_p=\di\alpha_{p-1}+\extd^\dagger\beta_{p+1}+\gamma_p\,.
}
The number of harmonic $p$-forms on $\M$ is equal to the dimension of the $p$-th de Rham cohomology group $\mathbf{H}_{\text{dR}}^p(\M)$.

 Here we obtain the Hodge dual of $\mathbb{T}_{\un{i}^p}$  $p$-forms in terms themselves which is useful to obtain the relations \eqref{phi psi relations duality}. This can be easily seen in form notation. First note that we have $\Delta \star=\star \Delta$. So $\star \mathbb{T}_{\un{i}^p}$ is an eigen-form of $\Delta$ with the same eigen-value as  $\mathbb{T}_{\un{i}^p}$
\begin{equation}
    \Delta\star\mathbb{T}^{(k)}_{\un{i}^p}=\star\Delta\mathbb{T}^{(k)}_{\un{i}^p}=\star\kk_p^2\mathbb{T}^{(k)}_{\un{i}^p}.
\end{equation}
Since we assumed that $\mathbb{T}_{\un{i}^p}^{(k)}$ is an co-exact form, then $\star \mathbb{T}_{\un{i}^p}^{(k)}$ is an exact $(n-p)$-form. So we can write, 
\begin{equation}
    \star \mathbb{T}_{\un{i}^p}^{(k)}=c(k,p) \di\mathbb{T}^{(k)}_{\un{i}^{n-p-1}}.
\end{equation}
Note that we also have $\Delta d=d \Delta$. So the consistency implies that $\kk_p^2=\kk_{n-p-1}^2$ which can be checked independently. The factor $c(k,p)$ can also be fixed by normalization condition,
\begin{equation}
    \langle \star\mathbb{T}_{\un{i}^p}^{(k)},\star\mathbb{T}_{\un{i}^p}^{(k)}\rangle=c^2 \langle \di\mathbb{T}_{\un{i}^{n-p-1}}^{(k)},\di\mathbb{T}_{\un{i}^{n-p-1}}^{(k)}\rangle=c^2\kk_{n-p-1}^2=c^2\kk_{p}^2=1.
\end{equation}
If $p=0$, the kernel of $\star d$ is the constant function (being a harmonic form) for which $k=0$ and it is dual to the volume form on the sphere.


\bibliographystyle{fullsort.bst}
 
\bibliography{review}

\end{document}